\documentclass[a4paper,11pt]{article}
\usepackage{jheppub} 

\usepackage[utf8]{inputenc}
\usepackage[colorlinks=true,citecolor=blue,linkcolor=blue]{hyperref}
\usepackage[normalem]{ulem}
\usepackage{amsmath,amssymb}
\usepackage{epsfig}
\usepackage{graphicx} 
\usepackage{placeins} 
\usepackage{url}
\usepackage{color}
\usepackage{dsfont}
\usepackage{slashed}
\usepackage{multirow}
\usepackage{enumerate}
\usepackage{placeins}
\usepackage[dvipsnames]{xcolor}
\usepackage{epstopdf}
\usepackage{soul}
\usepackage{tikz}
\usepackage[capitalise, english]{cleveref}
\usepackage{siunitx}
\usepackage{xspace}
\usepackage{subcaption}
\usepackage{array}
\usetikzlibrary{trees}
\usepackage{comment}
\usepackage{longtable}
\usetikzlibrary{decorations.pathmorphing}
\usetikzlibrary{decorations.markings}
\usetikzlibrary{calc,tikzmark,fit,shapes.geometric,matrix,decorations.markings,arrows.meta,decorations.pathmorphing,patterns,positioning,snakes}

\tikzset
  {midarrow/.style={decoration={markings,mark=at position 0.5 with
     {\arrow[thin,xshift=2pt]{Triangle[length=4pt,#1]}}},postaction={decorate}}
  }

\tikzset{
proton/.style = {circle, draw=black, thin, fill=black!20!white, minimum size=#1,
              inner sep=0pt, outer sep=0pt},
proton/.default = 6pt % size of the circle diameter 
}

\tikzset{
blob/.style = {circle, draw=black, thin, preaction={fill, black!20!white}, pattern=north east lines, minimum size=#1,
              inner sep=0pt, outer sep=0pt},
blob/.default = 6pt % size of the circle diameter 
}

\tikzset{
wc/.style = {circle, fill, minimum size=#1,
              inner sep=0pt, outer sep=0pt},
wc/.default = 4pt % size of the circle diameter 
}

\tikzset{
wcsq/.style = {rectangle, fill, red, minimum size=#1,
              inner sep=0pt, outer sep=0pt},
wcsq/.default = 8pt % size of the circle diameter 
}

\tikzset{vector/.style={decorate, decoration=snake}}
\tikzset{fermion/.style={decoration={markings, mark=at position 0.5 with {\arrow{>}}},postaction={decorate}}}
\newcommand\myshade{80}
\colorlet{mylinkcolor}{ForestGreen}
\colorlet{mycitecolor}{Red}
\colorlet{myurlcolor}{violet}

\hypersetup{
  linkcolor  = mylinkcolor!\myshade!black,
  citecolor  = mycitecolor!\myshade!black,
  urlcolor   = myurlcolor!\myshade!black,
  colorlinks = true
}
\definecolor{jblue}{RGB}{20,50,100}
\definecolor{npurple}{RGB} {153, 51, 204}
\definecolor{wred}{RGB}{217,0,56}
\definecolor{white}{RGB}{255,255,255}

\definecolor{korange}{RGB}{235, 80,  43}
\definecolor{korange2}{RGB}{245, 100,  63}
\definecolor{kyelloworange}{RGB}{255, 210,  110}
\definecolor{kyelloworange2}{RGB}{240, 170,  90}
\definecolor{kred}{RGB}{204,  102, 153}
\definecolor{kpurple}{RGB}{153,  61, 190}
\definecolor{kpurplelight}{RGB}{213,  161, 230}

\allowdisplaybreaks

\def\hyp{\mathsf{y}}

\newcommand{\eps}{\varepsilon}

\newcommand{\TeV}{{\,\rm TeV}}
\newcommand{\cC}{\mathcal{C}}
\newcommand{\cS}{\mathcal{S}}

\newcommand{\cO}{\mathcal{O}}

\usepackage{longtable}
\newcommand{\toprule}{}
\newcommand{\midrule}{}
\newcommand{\bottomrule}{}

\makeatletter 
\gdef\@fpheader{}
\makeatother
\begin{document}

%=============================================================================

\title{Accuracy complements energy: electroweak precision tests at Tera-Z}

\author[a]{Victor Maura,}
\emailAdd{victor.maura\_breick@kcl.ac.uk}

\author[b]{Ben A. Stefanek,}
\emailAdd{bstefan@ific.uv.es}

\author[a]{and Tevong You}
\emailAdd{tevong.you@kcl.ac.uk}

\affiliation[a]{Physics Department, King’s College London, Strand, London, WC2R 2LS, United Kingdom}

\affiliation[b]{Instituto de F\'isica Corpuscular (IFIC), Consejo Superior de Investigaciones \\Cient\'ificas (CSIC) and Universitat de Val\`{e}ncia (UV), 46980 Valencia, Spain}

\date{\today}

\preprint{KCL-PH-TH/2024-76}

%=============================================================================

\abstract{
A Tera-$Z$ factory, such as FCC-ee or CEPC, will have indirect sensitivity to heavy new physics up to the tens of TeV scale through higher-order loop contributions to precision measurements at the $Z$ pole. These indirect quantum effects may provide complementary, or even better, sensitivity to potential deviations from the Standard Model that are typically thought to best be constrained at leading order at higher energies above the $Z$ pole. We show in the SMEFT framework how accuracy complements energy for operators that modify the Higgs and gauge boson two- and three-point functions, leading to improved projected sensitivities for models such as the real singlet scalar, weakly interacting massive particles, and a custodial weak quadruplet. A thorough Tera-$Z$ programme may thus anticipate aspects of physics runs at higher energies and provide a wider scope of quantum exploration of the TeV scale than had previously been appreciated. 
}

\maketitle

%------------------------------------------------
\section{Introduction}
\label{sect:intro}
%------------------------------------------------

Studying the Higgs boson more precisely at a future Higgs factory has been identified as one of the main priorities for particle physics~\cite{CERN-ESU-015, osti_2368847}. A circular $e^+ e^-$ collider is arguably the best option for refining Higgs measurements to the sub-percent level and probing the Standard Model (SM) as generally as possible across all its intricately interconnected sectors~\cite{Blondel:2024mry}. With an extensive physics programme consisting of runs at the $Z$ pole, $WW$, $ZH$, and $t\bar{t}$ thresholds, it would not only guarantee results that would improve tremendously our picture of the fundamental building blocks of the universe, but also enable a far wider general exploration of the TeV scale than na\"{\i}vely expected. In particular, recent studies have highlighted how indirect quantum effects at the $Z$ pole, accessible by high precision measurements, can be a powerful and versatile probe of heavy new physics (NP)~\cite{Allwicher:2023aql,Allwicher:2023shc,Stefanek:2024kds,Allwicher:2024sso, Erdelyi:2024sls, Gargalionis:2024jaw}.  

At loop level, a Tera-$Z$ factory would have sensitivity to all Beyond the Standard Model (BSM) particles coupling linearly to the SM up to several tens of TeV in mass scale~\cite{Allwicher:2024sso, Gargalionis:2024jaw}~\footnote{The few exceptions for single-particle extensions can already be excluded by current experiments and at HL-LHC up to high scales~\cite{Davighi:2024syj}.}, even for unit couplings, and could probe new physics coupling dominantly to the third generation whose scale may still be as low as $\sim 1.5$ TeV without incurring any tension from stringent flavour constraints~\cite{Allwicher:2023aql,Allwicher:2023shc}. Furthermore, scenarios where the NP sector is custodially symmetric, as commonly assumed in composite Higgs models, will no longer evade electroweak precision tests due to the sensitivity of Tera-$Z$ to unavoidable SM custodial violations induced by renormalization group evolution (RGE)~\cite{Stefanek:2024kds}. Exploiting higher-order quantum effects thus extends the BSM physics case for FCC-ee~\cite{FCC:2018evy, Bernardi:2022hny} or CEPC~\cite{CEPCStudyGroup:2023quu} from intensively studied scenarios of light new physics, such as axions, dark photons, and heavy neutral leptons, to include a cornucopia of particles and theories above the weak scale that are also expected in many well-motivated BSM extensions. 

In this paper we further study the complementarity between high accuracy precision measurements on the $Z$ pole and measurements at higher energy runs that we collectively denote as ``above-pole". It is typically thought that the $Z$ pole, while strongly constraining the oblique $\hat{S}$ and $\hat{T}$ parameters~\cite{Peskin:1991sw}, is not particularly sensitive to modifications of Higgs couplings, pure gauge operators, or four-fermion contact interactions that are best constrained at higher energies. The LHC has also demonstrated how energy helps accuracy when compared to LEP constraints on the $\hat{W}$ and $\hat{Y}$ parameters~\cite{Farina:2016rws}. However, recent work has shown that hundreds of fermionic operators in the SM Effective Field Theory (SMEFT) can be efficiently probed on-pole at FCC-ee at Next-to-Leading-Order (NLO)~\cite{Allwicher:2023shc, Dawson:2022bxd, Bellafronte:2023amz}. Here we extend applications of the ``accuracy complements energy" principle by pointing out that the Higgs-only and gauge-only bosonic operators can also be constrained at the $Z$-pole with similar or better accuracy than above-pole measurements in higher energy runs, even for operators that are energy enhanced. 

Despite entering at NLO on the $Z$ pole, the higher statistics of a Tera-$Z$ factory can compensate for the loop suppression. Since the loop penalty leads to an NLO constraint that is a factor of $\sim 1/16\pi^2$ worse than the one at leading order, say in the $ZH$ run, we therefore need an increase in the number of events $N_Z$ at the $Z$ pole relative to $N_{ZH}$ at $ZH$ such that the statistical improvement $\sqrt{N_Z/N_{ZH}}$ satisfies
\begin{equation}
    \Delta^{NLO/LO}_{Z/ZH} \equiv \frac{1}{16\pi^2}\frac{\epsilon_Z}{\epsilon_{ZH}}\sqrt{\frac{N_Z}{N_{ZH}}} \gtrsim 1 \, ,
\end{equation}
where $\epsilon_{Z}$ and $\epsilon_{ZH}$ account for any reduction from the naive statistical expectation due to experimental and theoretical uncertainties~\cite{Bernardi:2022hny, Freitas:2019bre}, and we defined an improvement factor $\Delta^{NLO/LO}$. Assuming conservatively that $\epsilon_Z \sim 10^{-1}$ and $\epsilon_{ZH} \sim 1$, we see for $N_Z \sim 10^{12}$ and $N_{ZH} \sim 10^6$ that indeed $\Delta^{NLO/LO}_{Z/ZH}$ can be $O(1)$. We may therefore expect competitive sensitivity at NLO on the $Z$ pole with respect to above-pole constraints at leading order for any Wilson coefficient that both have access to. 

This argument also applies to new physics entering at leading order in both $Z$ pole and above-pole observables but with effects that grow with energy, as for the oblique $\hat{W}$ and $\hat{Y}$ parameters~\cite{Maksymyk:1993zm, Barbieri:2004qk}. An energy enhancement at higher energies, for example in $WW$ measurements, can be compensated for by higher statistical accuracy on-pole if  
\begin{equation}
    \Delta^{LO/LO}_{Z/WW} \equiv \frac{m_Z^2}{E_{WW}^2}\frac{\epsilon_Z}{\epsilon_{WW}}\sqrt{\frac{N_Z}{N_{WW}}} \gtrsim 1 \, ,
\end{equation}
where we defined the improvement factor $\Delta^{LO/LO}$ to include energy-enhanced terms. For $WW$ energies $E_{WW} \sim 200$ GeV, $\epsilon_Z\ \sim 10^{-1}$, $\epsilon_{WW} \sim 1$, $N_Z \sim 10^{12}$ and $N_{WW} \sim 10^8$, we obtain $\Delta^{LO/LO}_{Z/WW} = O(1)$. We note that $N_{WW} \approx N_{\bar f f} \approx 10^8$, so the same argument applies for energy enhancement in the $e^+ e^- \rightarrow \bar f f$ process. Thus, unlike at LEP (where $N_Z \sim 10^7$ and $N_{WW} \sim 10^5$), a Tera-Z factory is expected to be just as sensitive to the $\hat{W}$ and $\hat{Y}$ parameters on the $Z$ pole as in higher-energy $WW$ runs. Furthermore, it is important to combine all above-pole runs, as the number of fermion pairs produced falls off at roughly the same rate as the energy enhancement. This was recently shown explicitly by comparing $e^+ e^- \rightarrow \bar f f$ above-pole processes with $Z$-pole data in Ref.~\cite{Greljo:2024ytg}.

In our analysis we will confirm this expectation, finding that many operators typically thought to be mainly constrained at leading order in the higher-energy $WW$, $ZH$, and $t\bar{t}$ runs can actually receive better or comparable constraints at NLO on the Z-pole. Moreover, including on-pole data eliminates directions of limited sensitivity in a global fit which can remove degeneracies and improve constraints when fitting to several correlated operators at a time, as is necessary in most realistic ultraviolet (UV) models.

We indeed find this to be the case in several specific UV models that we study as examples. In the real scalar singlet and custodial quadruplet~\cite{Durieux:2022hbu} models, it is essential to take into account the correlated contributions to the Higgs self-coupling and Higgs self-energy when estimating the projected sensitivities in a global fit, even when there is a loop hierarchy between the two as for the weak quadruplet. Including these correlations weaken the bounds from above-pole measurements compared to considering only individual constraints; however, adding on-pole data then strengthens them further. For the real singlet scalar in particular, the on-pole information is crucial for potentially covering the entire region of parameter space in which the scalar can be a so-called ``loryon"~\cite{Banta:2021dek}, thus closing the last remaining open window in being able to conclusively answer whether any other particles outside the SM get most of their mass from the Higgs~\cite{Crawford:2024nun}. We also consider the case of Weakly Interacting Massive Particles (WIMPs) and show that electroweak precision can significantly improve on the picture after HL-LHC for this elusive class of BSM extensions.  

In Sec.~\ref{sec:setupobs}, we begin by describing the observables used in our analysis and the construction of our likelihood. Sec.~\ref{sec:EFTanalysis} introduces the SMEFT framework and conventions we adopt before studying how accuracy complements energy for the Higgs coupling modifications, pure gauge operators, and four-fermion contact interactions. Sec.~\ref{sec:models} applies this to specific example UV models: the real singlet scalar, WIMPs, and a custodial weak quadruplet scalar, before concluding with a discussion in Sec.~\ref{sec:conclusions}. 

\section{Setup and Observables}
\label{sec:setupobs}

The measurements used in our analysis are classified as on-pole for $Z$- and $W$-pole observables, and as above-pole observables for everything else. For \emph{on-pole} data, we use the following $Z$-pole observables,
\begin{equation*}
    O_\text{$Z$-pole} = \left\{ \Gamma_Z \, , \sigma_\text{had} \, , R_l, A_\text{FB}^{0,l} \, , R_b \, , R_c \, , A_b^\text{FB} \, , A_c^\text{FB} \, , A_l \, , A_b \, , A_c \, , A_s \, \right\} \, ,
\end{equation*}
where $l=e,\mu,\tau$, and the observables are defined as
\begin{align*}
\Gamma_Z &\equiv \sum_f \Gamma\left(Z \to f \bar{f}\right) \, , \quad 
\sigma_\text{had} \equiv \frac{12\pi}{m_Z^2}\frac{\Gamma(Z\to e^+ e^-)\Gamma(Z \to q\bar{q})}{\Gamma_Z^2} \, , \\ 
R_l &\equiv \frac{\sum_q \Gamma(Z\to q\bar{q})}{\Gamma(Z\to l^+l^-)} \, , \quad
A_\text{FB}^{0,l} \equiv \frac{3}{4}A_e A_l \, , \\
R_q &\equiv \frac{\Gamma(Z\to q \bar{q})}{\sum_{q_i} \Gamma(Z \to q_i \bar{q}_i)} \, , \quad
A_f \equiv \frac{\Gamma(Z\to f_L^+ f_L^-) - \Gamma(Z \to f_R^+ f_R^-)}{\Gamma(Z\to f^+ f^-)} \, ,
\end{align*}
and we include the following selection of $W$-pole observables, 
\begin{align*}
    O_\text{$W$-pole} = \left\{ m_W \, , \Gamma_W \, \right\} \, .
\end{align*}
For these observables, we use the expressions calculated at NLO in the SMEFT with \{$G_F$, $m_Z$, $\alpha$\} as input parameters in Ref.~\cite{Bellafronte:2023amz}. The relative uncertainties and projected error reductions for the $Z$- and $W$-pole observables are given in Table~\ref{tab:FCCeePROJZpole}. The $W$ pole uncertainties are from Ref.~\cite{Bernardi:2022hny}. The $Z$ pole experimental uncertainties are taken from Ref.~\cite{DeBlas:2019qco}. We neglect theoretical uncertainties, assuming the ``TH3" scenario of Ref.~\cite{Blondel:2018mad}. The sole exception to this is $\Gamma_Z$~\cite{Bernardi:2022hny, Freitas:2019bre}, which will likely remain theory limited with an error of $O(100)$ keV, which we use to determine the $\Gamma_Z$ improvement factor in Table~\ref{tab:FCCeePROJZpole}.

\begin{table}[t!]
 \centering
\renewcommand{\arraystretch}{1.} 
 \begin{tabular}{|c|c|c|c|}
\hline
 & Current Rel.  & FCC-ee Rel.  & Proj. Error \\
Observable & Error ($10^{-3}$) & Error ($10^{-3}$) & Reduction \\
\hline
\hline 
$ \mathrm{  \Gamma_Z } $ &  0.92& 0.04 & 23 \\
\hline
$ \sigma_{\rm had}^0$ &  0.78& 0.11 & 7.4  \\
\hline
$ R_b $ &  3.06 & 0.3 & 10.2    \\
\hline
$ R_c $ &  17.4 & 1.5 & 11.6    \\
\hline
$ A_{\rm FB}^{0,b} $ &  15.5 & 1 & 15.5    \\
\hline
$ A_{\rm FB}^{0,c} $ &  47.5 & 3.08 & 15.4    \\
\hline
$ A_b $ &  21.4 & 3 & 7.13    \\
\hline
$ A_c $ &  40.4 & 8 & 5.05    \\
\hline
$ R_e $ &  2.41 & 0.3 & 8.03    \\
\hline
$ R_\mu $ &  1.59 & 0.05 & 31.8    \\
\hline
$ R_\tau $ &  2.17 & 0.1 & 21.7    \\
\hline
$ A_{\rm FB}^{0,e} $ &  154 & 5 & 30.8    \\
\hline
$ A_{\rm FB}^{0,\mu} $ &  80.1 & 3 & 26.7    \\
\hline
$ A_{\rm FB}^{0,\tau} $ &  104.8 & 5 & 21    \\
\hline
$ A_e^{**}$ &  12.5 & 0.13 & 95    \\
\hline
$ A_\mu^{**} $ &  102 & 0.15 & 680    \\
\hline
$ A_\tau^{**} $ &  102 & 0.3 & 340    \\
\hline
$ \mathrm{  \Gamma_W} $ & 20.1 & 0.59& 34 \\
\hline
$ \mathrm{  m_W} $ & 0.15 & 0.004& 38\\
\hline
 \end{tabular}
 \caption{Projected FCC-ee improvement for $Z$-pole observables and $W$ pole observables from \cite{Bernardi:2022hny, DeBlas:2019qco}. The $A_\ell^{**}$ are from lepton
polarization and LR asymmetry measurements at SLC. For $ A_e^{**}$, we use the projection in~\cite{deBlas:2022ofj}. }
 \label{tab:FCCeePROJZpole}
 \end{table}

On the other hand, the \emph{above-pole} processes considered here are
\begin{equation*}
    O_\text{above-pole} = \left\{ \sigma\left(e^+ e^- \to W^+ W^-\right), \, \sigma\left(e^+ e^- \to ZH\right), \, 
    \sigma\left(e^+ e^- \to f\bar{f}\right) \,
    \right\} .
\end{equation*}
For the $ZH$ production cross section, we use the results from Ref.~\cite{Asteriadis:2024xts}, which have been computed at NLO in the SMEFT. We use our own leading order (LO) calculation for the $WW$ process, given in Appendix~\ref{app:WW}. For both calculations, the \{$G_F$, $m_W$, $m_Z$\} input scheme is adopted.\footnote{The choice of input parameter scheme modifies the predictions in two ways: it modifies the numerical value of the SM parameters used in the calculations and shifts the observable by a linear combination of $\mathcal{C}_i$. Using the results of Ref. \cite{Biekotter:2025nln}, we have verified that the former is numerically very small. The latter effect is generally larger, but is confined to the coefficients $C_{HD}, C_{HWB},[ C_{Hl}^{(3)}]_{11},[ C_{Hl}^{(3)}]_{22},[C_{ll}]_{1221}$~\cite{Biekotter:2023xle}. These are constrained far more precisely at the Z-pole, so we do not expect the scheme difference has any impact on the sensitivity where these operators are concerned.}

Regarding the $e^+ e^- \to f \bar{f}$  process, we use the results of the recent dedicated flavor-tagging analysis for FCC-ee performed in Ref.~\cite{Greljo:2024ytg}. They use the cross-section ratio observables
\begin{equation*}
R_{f} = \frac{\sigma(e^+e^-\rightarrow f\bar{f})}{\sum_{q}\sigma(e^+e^-\rightarrow q\bar{q})}
\end{equation*}
where the sum runs over all the light quarks. Our predictions are implemented following Appendix B of Ref.~\cite{Greljo:2024ytg}. 

We first express all observables as functions of the Wilson coefficients at the scale of the observable, which we take to be $\mu = m_Z$ for the $Z/W$-pole and $\mu = \sqrt{s} = 163, 240, 365$ GeV for  observables above the pole. We then perform the RGE up to a high scale $\Lambda = 1$ TeV to express all observables as a function of the high scale Wilson coefficients, which are then used to construct the likelihood. The fit is then performed via a simple $\chi^2$ procedure as in Ref.~\cite{Allwicher:2023aql,Allwicher:2023shc,Stefanek:2024kds}. The leading order $Z/W$-pole part of this likelihood has previously been applied and validated in Refs.~\cite{Allwicher:2023aql,Davighi:2023evx,Allwicher:2023shc,Stefanek:2024kds}, and has also been cross-checked here with an independent implementation in {\tt fitmaker}~\cite{Ellis:2020unq}. Finally, we note that the expected sensitivity for all observables has been appropriately rescaled to match the luminosity figures given in the FCC Feasibility Study Report~\cite{Benedikt:2928193}.

\subsection*{Effective field theory framework}
Effective field theory is an appropriate framework for capturing the indirect effects of heavy new physics in a relatively model-independent way. We assume here a linear realisation of the SM $SU(2) \times U(1)$ gauge symmetry with the Higgs as part of an $SU(2)$ doublet. Working with the SMEFT at the dimension-6 level in the Warsaw basis, we may write the relevant dimension-6 Lagrangian terms as 
\begin{equation}
    \mathcal{L}_{\rm SMEFT} \supset \sum_{i} \frac{C_{i}}{\Lambda^2} Q_{i}^{\rm Warsaw} \, ,
\end{equation}
where $Q_i$ are dimension-6 operators and $C_i$ are the dimensionless Wilson coefficients generated by heavy new physics at a scale $\Lambda$. We will take $\Lambda = 1$ TeV unless otherwise stated. 

In some cases, it will prove convenient to use the SILH basis~\cite{Giudice:2007fh,Contino:2013kra}, such as to characterise pure gauge operators. To avoid conflict with Warsaw basis operators and Wilson coefficients, when working in the SILH basis we adopt the following notation
\begin{equation}
    \mathcal{L}_{\rm SMEFT} \supset \sum_{i} \frac{S_{i}}{\Lambda^2} \mathcal{O}_{i}^{\rm SILH} \, ,
\end{equation}
where the operators $\mathcal{O}_{i}^{\rm SILH}$ will always be explicitly defined in the text for convenience.

Finally, for more concise expressions we will sometimes employ dimensionful Wilson coefficients defined as $\mathcal{C}_i \equiv C_i/\Lambda^2$ and $\mathcal{S}_i \equiv S_i/\Lambda^2$.

% %------------------------------------------------
\section{Accuracy complements energy: an EFT analysis}
\label{sec:EFTanalysis}

The SMEFT language provides a systematic way of exploring modifications from heavy BSM physics that are correlated across different measurements at various energies (for some recent global fits, see e.g. Refs.~\cite{Ellis:2020unq, Bissmann:2020mfi, Ethier:2021bye, Grunwald:2023nli, Garosi:2023yxg, Allwicher:2023shc, Bartocci:2023nvp, Celada:2024mcf, deBlas:2022ofj, Bartocci:2024fmm}). Our approach is to focus on operators contributing to measurements above the $Z$ pole (denoted above-pole), some of which are energy enhanced, and that also enter the $Z$-pole at one higher loop order.

\begin{figure}[t]
    %\hspace{-7.5mm}
    \includegraphics[width=1\textwidth]{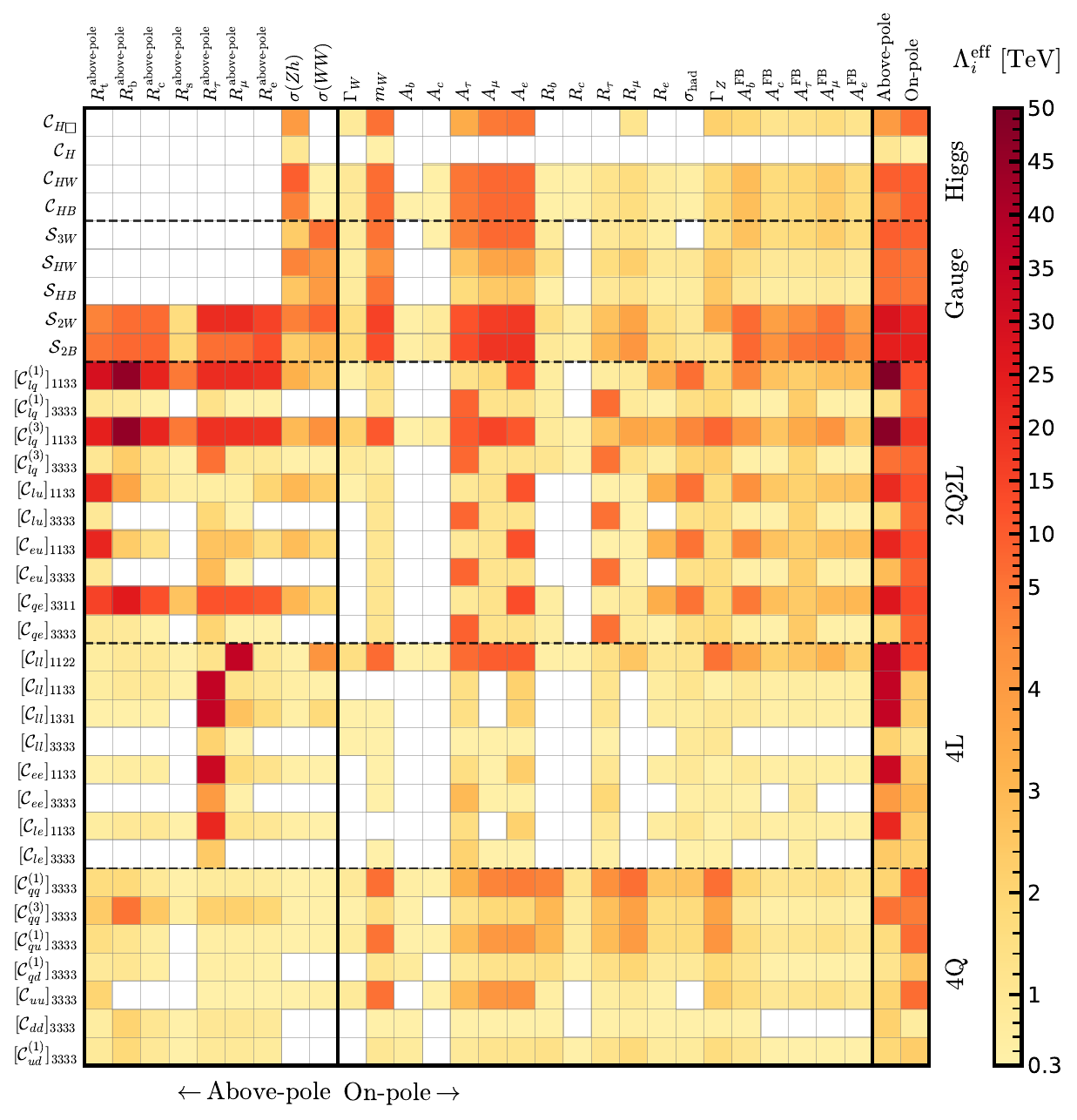}
    \caption{Projected sensitivity at FCC-ee of measurements on the $Z$ pole and above-pole at higher energy runs for Higgs, gauge, and four-fermion quark/lepton operators. The colour gradient denotes the bound on the scale $\Lambda$ in TeV for unit Wilson coefficient. }
    \label{fig:heatMap}
\end{figure}

\cref{fig:heatMap} summarises the complementarity between on-pole and above-pole constraints. The effects of heavy new physics are parametrised by the Wilson coefficients $\mathcal{C}_i \equiv C_i/\Lambda^2$ of dimension-6 operators $Q_i$ in the Warsaw basis of the SMEFT~\cite{Grzadkowski:2010es}, while the projected sensitivity at FCC-ee to the effective scale of new physics is represented by the colour-coded heatmap gradient. The operators shown are grouped by Higgs, gauge, and various four-fermion operators, and are typically thought to best be constrained at leading order above the pole at the $WW$, $ZH$, and $\bar t t$ higher-energy runs. Darker shading indicates a stronger constraint, and we see in many cases the on-pole constraint dominating or giving a similar order of magnitude sensitivity to the corresponding above-pole constraint. This is indeed the case for the $\hat{W}$ and $\hat{Y}$ oblique parameters that map on to the following  operators in the SILH basis~\cite{Giudice:2007fh,Contino:2013kra}
\begin{equation}
\mathcal{L}_{\rm SMEFT} \supset -\frac{\mathcal{S}_{2W}}{2}(D^{\mu}W_{\mu \nu}^I) (D_{\rho}W^{I \rho \nu}) -\frac{\mathcal{S}_{2B}}{2}(\partial^{\mu}B_{\mu \nu}) (\partial_{\rho}B^{ \rho \nu}) \,.
\end{equation}
For other operators the above-pole constraint can win due to energy enhancement effects, though the benefit of entering at leading order vanishes when considering four-fermion operators involving second and third generation fermions. Even when the sensitivity is worse, the complementarity of the on-pole constraints helps close flat directions in global fits and leads to better marginalised constraints or bounds on specific models. 

We now illustrate in more detail how accuracy complements energy with Higgs sector modifications before considering pure gauge operators and then summarising the recent four-fermion results. 

\subsection{Higgs coupling modifications}
\begin{figure}[t!]
    \centering
    \begin{tabular}{c@{\hskip 1cm}c@{\hskip 1cm}c}
        \begin{tikzpicture}[thick,>=stealth,baseline=-0.5ex]
            \draw[dashed] (-2,0) -- (-0.7,0);
            \draw[dashed] (-0.7,0) -- (0.7,0);
            \node[wcsq] at (-0.7,0) {};
             %%%%%%%%%%Labels
            % \node[left] at (-1.7,0) {$Z$};
            % \node[right] at (1.7,0) {$Z$};
            % \node[above] at (0,0.7) {$t$};
            % \node[below] at (0,-0.7) {$t$};
            % \node[right] at (-0.7,0.0) {$\delta g_{R33}^{Zu}$};
        \end{tikzpicture}
        &
        \begin{tikzpicture}[thick,>=stealth,baseline=-0.5ex]
            \draw[vector] (-2.5,0.2) -- (-0.7,0.2);
            \draw[vector] (-0.7,0.2) -- (0.15,1.4);
            % \draw[dashed] (-1.55,0.13) -- (-0.5,-0.3);
            % \draw[dashed] (-0.25,0.7) -- (-0.5,-0.3);
            \draw[dashed] (-0.7,0.2) -- (0.25,-0.6);
            \node[wcsq] at (-0.25,-0.19) {};
             %%%%%%%%%%Labels
            % \node[left] at (-1.7,0) {$Z$};
            % \node[right] at (1.7,0) {$Z$};
            % \node[above] at (0,0.7) {$t$};
            % \node[below] at (0,-0.7) {$t$};
            % \node[right] at (-0.7,0.0) {$\delta g_{R33}^{Zu}$};
        \end{tikzpicture}
        & 
        \begin{tikzpicture}[thick,>=stealth,baseline=-0.5ex]
            \draw[vector] (-2,0) -- (-0.7,0);
            \draw[dashed] (-0.7,0) arc (180:0:0.7);
            % \draw[dashed] (0,0.71) -- (0,-0.85);
            \draw[vector] (0.7,0) arc (0:-180:0.7);
            \draw[vector] (0.7,0) -- (2,0);
            \node[wcsq] at (0,0.71) {};
             %%%%%%%%%%Labels
            % \node[left] at (-1.7,0) {$Z$};
            % \node[right] at (1.7,0) {$Z$};
            % \node[above] at (0,0.7) {$t$};
            % \node[below] at (0,-0.7) {$t$};
            % \node[right] at (-0.7,0.0) {$\delta g_{R33}^{Zu}$};
        \end{tikzpicture}
        \\
        (a) Higgs self-energy &  (b) $e^+ e^- \rightarrow ZH$ &  (c) $Z$-pole oblique params.
    \end{tabular}
    \caption{New physics affecting the Higgs self-energy in (a), contributing to Higgs observables at LO in (b), and $Z$-pole observables at NLO in (c). Red squares indicate SMEFT contributions to the Higgs propagator.}
    \label{fig:CHBoxdiagrams}
\end{figure}
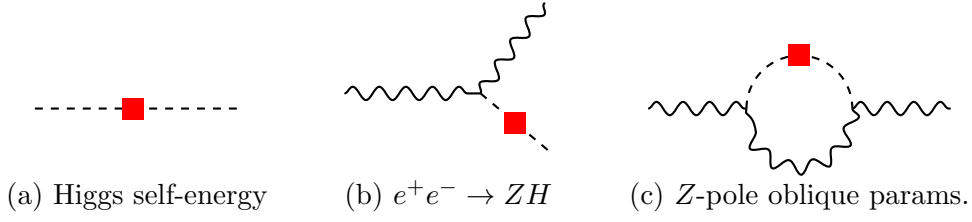
\begin{figure}[t!]
    \centering
    \begin{tabular}{c@{\hskip 1cm}c@{\hskip 1cm}c}
        \begin{tikzpicture}[thick,>=stealth,baseline=-0.5ex]
            \draw[dashed] (-2,0) -- (-0.7,0);
            \draw[dashed] (-0.7,0) -- (0,1);
            \draw[dashed] (-0.7,0) -- (0,-1);
            \node[wcsq] at (-0.7,0) {};
             %%%%%%%%%%Labels
            % \node[left] at (-1.7,0) {$Z$};
            % \node[right] at (1.7,0) {$Z$};
            % \node[above] at (0,0.7) {$t$};
            % \node[below] at (0,-0.7) {$t$};
            % \node[right] at (-0.7,0.0) {$\delta g_{R33}^{Zu}$};
        \end{tikzpicture}
        & 
        \begin{tikzpicture}[thick,>=stealth,baseline=-0.5ex]
            \draw[vector] (-2.5,0.2) -- (-0.7,0.2);
            \draw[vector] (-0.7,0.2) -- (0.15,1.4);
            \draw[dashed] (-1.55,0.13) -- (-0.5,-0.3);
            \draw[dashed] (-0.25,0.7) -- (-0.5,-0.3);
            \draw[dashed] (-0.5,-0.3) -- (0.25,-0.8);
            \node[wcsq] at (-0.5,-0.3) {};
             %%%%%%%%%%Labels
            % \node[left] at (-1.7,0) {$Z$};
            % \node[right] at (1.7,0) {$Z$};
            % \node[above] at (0,0.7) {$t$};
            % \node[below] at (0,-0.7) {$t$};
            % \node[right] at (-0.7,0.0) {$\delta g_{R33}^{Zu}$};
        \end{tikzpicture}
        & 
        \begin{tikzpicture}[thick,>=stealth,baseline=-0.5ex]
            \draw[vector] (-2,0) -- (-0.7,0);
            \draw[dashed] (-0.7,0) arc (180:0:0.7);
            \draw[dashed] (0,0.71) -- (0,-0.85);
            \draw[vector] (0.7,0) arc (0:-180:0.7);
            \draw[vector] (0.7,0) -- (2,0);
            \node[wcsq] at (0,0.71) {};
             %%%%%%%%%%Labels
            % \node[left] at (-1.7,0) {$Z$};
            % \node[right] at (1.7,0) {$Z$};
            % \node[above] at (0,0.7) {$t$};
            % \node[below] at (0,-0.7) {$t$};
            % \node[right] at (-0.7,0.0) {$\delta g_{R33}^{Zu}$};
        \end{tikzpicture}
        \\
        (a) Higgs self-coupling &  (b) $e^+ e^- \rightarrow ZH$ &  (c) $Z$-pole oblique params.
    \end{tabular}
    \caption{New physics affecting the Higgs self-coupling in (a), contributing to Higgs observables at NLO in (b), and $Z$-pole observables at NNLO in (c). Red squares indicate SMEFT contributions to the vertex.}
    \label{fig:CHdiagrams}
\end{figure}
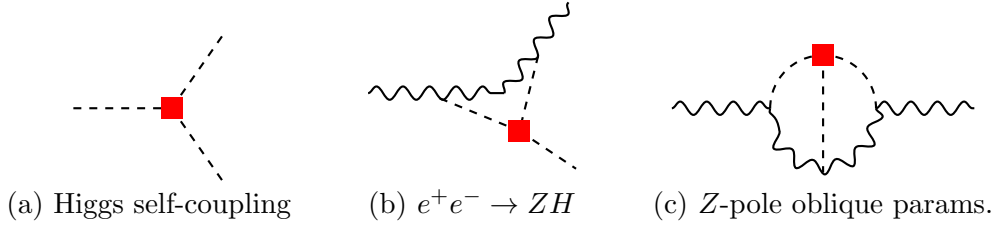
\begin{figure}[t!]
    \centering
    \begin{tabular}{c@{\hskip 1cm}c@{\hskip 1cm}c}
        \begin{tikzpicture}[thick,>=stealth,baseline=-0.5ex]
            \draw[vector] (-2,0) -- (-0.7,0);
            \draw[dashed] (-0.7,0) -- (0,1);
            \draw[vector] (-0.7,0) -- (0,-1);
            \node[wcsq] at (-0.7,0) {};
             %%%%%%%%%%Labels
            % \node[left] at (-1.7,0) {$Z$};
            % \node[right] at (1.7,0) {$Z$};
            % \node[above] at (0,0.7) {$t$};
            % \node[below] at (0,-0.7) {$t$};
            % \node[right] at (-0.7,0.0) {$\delta g_{R33}^{Zu}$};
        \end{tikzpicture}
        & \hspace{2cm}
        \begin{tikzpicture}[thick,>=stealth,baseline=-0.5ex]
            \draw[vector] (-2,0) -- (-0.7,0);
            \draw[dashed] (-0.7,0) arc (180:0:0.7);
            \draw[vector] (0.7,0) arc (0:-180:0.7);
            \draw[vector] (0.7,0) -- (2,0);
            \node[wcsq] at (-0.7,0) {};
             %%%%%%%%%%Labels
            % \node[left] at (-1.7,0) {$Z$};
            % \node[right] at (1.7,0) {$Z$};
            % \node[above] at (0,0.7) {$t$};
            % \node[below] at (0,-0.7) {$t$};
            % \node[right] at (-0.7,0.0) {$\delta g_{R33}^{Zu}$};
        \end{tikzpicture}
        \\
        (a) $hVV$ couplings & \hspace{2cm} (b) $Z$-pole oblique params.
    \end{tabular}
    \caption{New physics affecting the $hVV$ coupling in (a),  NLO contribution of $hVV$ coupling modifications to $Z$-pole observables. Red squares indicate SMEFT contributions to the vertex.}
    \label{fig:hVVdiagrams}
\end{figure}
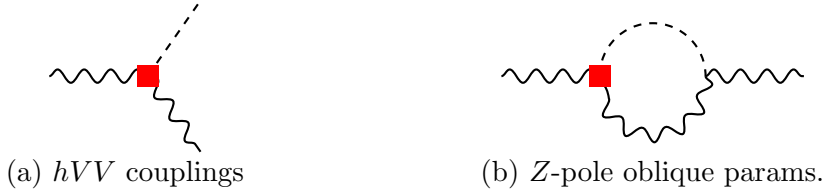

A Higgs factory will be sensitive at leading order to new physics modifying the Higgs self-energy in the process $e^+ e^- \to ZH$, as illustrated by the propagator correction in Fig.~\ref{fig:CHBoxdiagrams}; we also see in Fig.~\ref{fig:CHBoxdiagrams} that it necessarily also enters at NLO in the oblique parameters on the $Z$ pole. Similarly, Fig.~\ref{fig:CHdiagrams} shows that a vertex correction to the Higgs self-coupling will enter at NLO in $e^+ e^- \to ZH$ and at NNLO on the $Z$ pole, while Fig.~\ref{fig:hVVdiagrams} shows a LO $hVV$ vertex correction modifying the $Z$ pole oblique parameter at NLO. As argued previously, we expect the suppression from an extra loop factor on the $Z$ pole to be compensated by increased statistics, such that the on-pole vs. above-pole improvement factor on constraints from all these Higgs coupling modifications could be $\Delta^{NLO/LO}_{Z/ZH} \sim \mathcal{O}(1)$ for both the ratios NLO/LO and NNLO/NLO.  

The cross-section $\sigma(e^+ e^- \to ZH)$ in the SMEFT is sensitive to 3 (4) dimension-6 operators at LO (NLO) that can modify the Higgs couplings~\cite{Asteriadis:2024xts}, 
\begin{align}
Q_{H\Box} & = (H^\dagger H)\Box(H^\dagger H) \,,  \\
Q_{H} & = (H^\dagger H)^3 \,, \\
Q_{HW} &= (H^\dagger H) W^I_{\mu\nu}W^{I\,\mu\nu} \,, \\
Q_{HB} &= (H^\dagger H) B_{\mu\nu}B^{\mu\nu} \,.
\end{align}
The operators $Q_{H\Box}$, $Q_{HW}$, and $Q_{HB}$ contribute to the cross-section at leading order, whereas $Q_H$ first appears at NLO as a finite 1-loop contribution~\cite{McCullough:2013rea}. 

These operators then enter in $Z/W$-pole observables at NLO and NNLO, respectively. As can be seen from the 1-loop RG equations, $Q_{H\Box}$, $Q_{HW}$, and $Q_{HB}$ run into the most sensitive $Z$-pole Wilson coefficients~\cite{Alonso:2013hga}
\begin{align}
\dot{\cC}_{HD} &\supset \frac{80}{3} g_1^2 \hyp_h^2 \cC_{H\Box} \label{eq:CHboxRGE}\,, \\
[\dot{\cC}_{Hl}^{(3)}]_{pr} &\supset \frac{g_2^2}{6} \cC_{H\Box} \delta_{pr} \,, \label{eq:CHl3RGE} \\
\dot{\cC}_{HWB} &\supset 2g_1 g_2 (\cC_{HW} + \cC_{HB}) +3g_1 g_2^2 \cC_W\,.
\label{eq:HWBrge}
\end{align}
On the other hand, $\cC_H$ only enters the $Z/W$ pole observables as a finite 2-loop contribution that can be captured as a modification to the oblique $\hat S$ and $\hat T$ parameters~\cite{vanderBij:1985ww,Kribs:2017znd,Degrassi:2017ucl},
\begin{align}
10^6 \hat{S} &= -0.71 C_H - 0.25 C_H^2 \,, \nonumber \\
10^6 \hat{T} &= 1.2 C_H + 0.36 C_H^2 \,.
\label{eq:CH2loop}
\end{align}
We implement these contributions to $\hat{S}$ and $\hat{T}$ by their equivalent shifts to $\cC_{HWB}$ and $\cC_{HD}$, respectively. 

\begin{figure}[t]
    \centering
    \hspace{-15mm}
    \includegraphics[scale=0.8775]{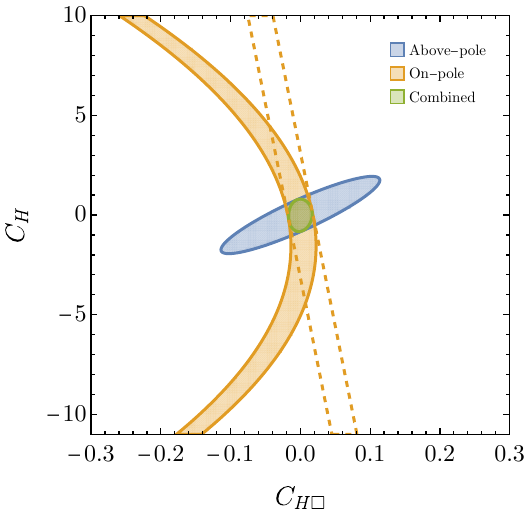} \hspace{2.5mm}
    \includegraphics[scale=0.88]{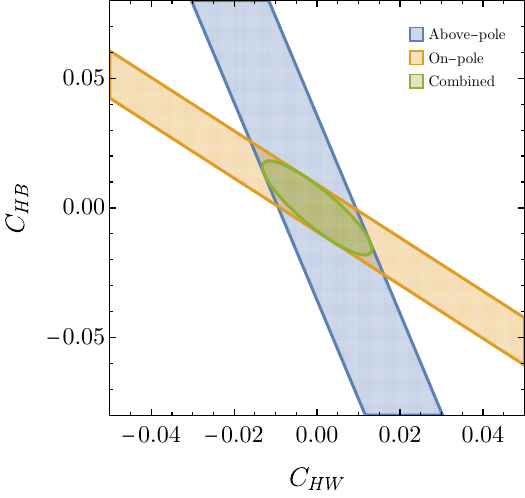}
    \caption{FCC-ee projected sensitivities at 68\% CL to Higgs coupling modifications on and off the $Z/W$ pole. Left: SMEFT operators that modify the Higgs 2- and 3-point functions. The dashed region is obtained by linearizing~\cref{eq:CH2loop}. Right: SMEFT operators giving $hVV$ coupling modifications. The Wilson coefficients are renormalised at $\Lambda = 1$ TeV.}
    \label{fig:higgsPlot}
\end{figure}

\subsubsection{Higgs self-energy and self-coupling}

The left panel of~\cref{fig:higgsPlot} shows the on- and above-pole 68\% CL projected sensitivities at FCC-ee in the $C_{H\Box}$ vs. $C_H$ plane, that modify the Higgs 2-point and 3-point functions respectively. The blue-shaded ellipse corresponds to the above-pole constraints (at LO for $C_{H\Box}$ and NLO for $C_H$) while the orange-shaded region denotes the on-pole bounds (at NLO for $C_{H\Box}$ and NNLO for $C_H$). We note that an otherwise flat direction for the above-pole observables is broken by measuring $\sigma(e^+e^- \to ZH)$ at 240 and 365 GeV. The on-pole observables then further constrain an almost orthogonal direction, leading to a clear complementarity in the combination shown by the green ellipse. While in general the region probed by on-pole observables depends on the inclusion of the quadratic terms in~\cref{eq:CH2loop}, the combination with above-pole $ZH$ data clearly limits $C_H$ to the linear regime represented by the region between dashed orange lines.

Constraining the Higgs self-coupling indirectly at NLO in $ZH$ was first proposed in Ref.~\cite{McCullough:2013rea}, and a subsequent study found that accounting for deviations in other SM couplings could still lead to a competitive determination at future lepton colliders~\cite{DiVita:2017vrr}. As we have emphasised here, however, it is important to include the full set of SMEFT effects at NLO in $ZH$ and NNLO on the $Z$ pole to account for any correlations in extracting bounds from a global fit. We leave such a study to future work.    

\subsubsection{Higgs-vector-vector couplings}

We now turn to the right panel of~\cref{fig:higgsPlot} which shows the on- and above-pole 68\% CL projected sensitivities at FCC-ee in the $C_{HW}$ vs. $C_{HB}$ plane that modify the 3-point $hVV$ vertex. The complementarity is again evident between the on-pole constraints in the orange-shaded region and the above-pole blue-shaded regions. Both on- and above-pole observables have flat directions that are lifted by the combination to give the combined green elliptical region.

\subsubsection{Higgs sector combined fit}

We report here a combined on- and above-pole likelihood for $C_H,C_{H\Box},C_{HW},C_{HB}$ from a 4-parameter Gaussian fit. The 1-$\sigma$ intervals for the Wilson coefficients renormalised at 1 TeV are
\begin{equation}
\left(
\begin{array}{cc}
 C_H  \\
 C_{H\Box} \\
 C_{HW} \\
 C_{HB} \\
\end{array}
\right)= \pm
\left(
\begin{array}{cc}
0.932 \\
0.052 \\
0.015 \\
0.031 \\
\end{array}
\right) \,,
\end{equation}
and the correlation matrix is
\begin{equation}
\left(
\begin{array}{cccc}
 1 & 0.174 & 0.772 & -0.612 \\
 0.174 & 1 & -0.122 & -0.786 \\
 0.772 & -0.122 & 1 & -0.489 \\
 -0.612 & -0.786 & -0.489 & 1 \\
\end{array}
\right)\,.
\end{equation}
We note that the combination of on-pole and above-pole data here enables a sensible closed fit to be obtained for this 4-parameter subset of Higgs coupling operators.

\subsection{Pure gauge operators}
\label{sec:PureGauge}

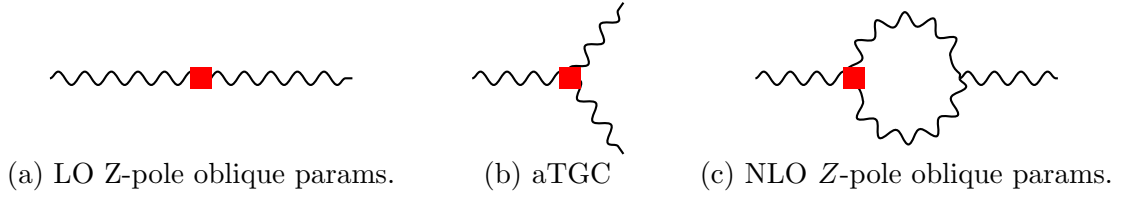
\begin{figure}[t!]
    \centering
        \begin{tabular}{c@{\hskip 1cm}c@{\hskip 1cm}c}
            \begin{tikzpicture}[thick,>=stealth,baseline=-0.5ex]
            \draw[vector] (-2,0) -- (2,0);
            \node[wcsq] at (0,0) {};
            %%%%%%%%%%Labels
            % \node[left] at (-1.7,0) {$Z$};
            % \node[right] at (1.7,0) {$Z$};
            % \node[above] at (0,0.7) {$t$};
            % \node[below] at (0,-0.7) {$t$};
            % \node[right] at (-0.7,0.0) {$\delta g_{R33}^{Zu}$};
        \end{tikzpicture}
        &
        \begin{tikzpicture}[thick,>=stealth,baseline=-0.5ex]
            \draw[vector] (-2,0) -- (-0.7,0);
            \draw[vector] (-0.7,0) -- (0,1);
            \draw[vector] (-0.7,0) -- (0,-1);
            \node[wcsq] at (-0.7,0) {};
             %%%%%%%%%%Labels
            % \node[left] at (-1.7,0) {$Z$};
            % \node[right] at (1.7,0) {$Z$};
            % \node[above] at (0,0.7) {$t$};
            % \node[below] at (0,-0.7) {$t$};
            % \node[right] at (-0.7,0.0) {$\delta g_{R33}^{Zu}$};
        \end{tikzpicture}
        &
        \begin{tikzpicture}[thick,>=stealth,baseline=-0.5ex]
            \draw[vector] (-2,0) -- (-0.7,0);
            \draw[vector] (-0.7,0) arc (180:0:0.7);
            \draw[vector] (0.7,0) arc (0:-180:0.7);
            \draw[vector] (0.7,0) -- (2,0);
            \node[wcsq] at (-0.7,0) {};
             %%%%%%%%%%Labels
            % \node[left] at (-1.7,0) {$Z$};
            % \node[right] at (1.7,0) {$Z$};
            % \node[above] at (0,0.7) {$t$};
            % \node[below] at (0,-0.7) {$t$};
            % \node[right] at (-0.7,0.0) {$\delta g_{R33}^{Zu}$};
        \end{tikzpicture}
        \\
        (a) LO Z-pole oblique params. & (b) aTGC & (c) NLO $Z$-pole oblique params.
    \end{tabular}
    \caption{LO new physics modifications of gauge 2- and 3-point functions (a+b) and NLO contribution of anomalous triple gauge couplings (aTGC) to $Z$-pole observables (c). Red squares indicate SMEFT contributions.}
    \label{fig:aTGCdiagrams}
\end{figure}

We move now to the gauge sector.
Fig.~\ref{fig:aTGCdiagrams} shows new physics modifications at leading order in the 2-point functions for the oblique parameters and in the 3-point vertex for anomalous Triple-Gauge-Couplings (aTGCs) that are constrained at LO by $e^+ e^- \rightarrow WW$ at higher energies. However, we see that the aTGC vertex modification also enters at NLO in the $Z$-pole oblique parameters, thus leading us to apply our accuracy complements energy argument here. 

Of the 4 oblique parameters $\hat{S}$, $\hat{T}$, $\hat{W}$, and $\hat{Y}$ that correspond to dimension-6 operators, it is thought that only $\hat{S}$ and $\hat{T}$ are strongly constrained on the $Z$ pole while $\hat{W}$ and $\hat{Y}$ are best constrained at higher energies due to their energy enhancement. However, as argued in the introduction, unlike at LEP this is no longer expected to be the case at FCC-ee, where the statistical improvement in accuracy at the $Z$ pole can compensate for the energy enhancement of above-pole processes despite both being at leading order. For example, in $WW$ processes this is due to the relative improvement in the number of $WW$ and $Z$ boson pairs in going from LEP to FCC-ee, $\left( N_Z/N_{WW} \right)_{FCC} \sim 10^5 \gg \left( N_Z/N_{WW} \right)_{LEP} \sim 10^2$.

\subsubsection{Oblique parameters $\hat{S}$, $\hat{T}$, $\hat{W}$, $\hat{Y}$}
\label{sec:ObParams}
We begin by considering the oblique parameters $\hat{S}$, $\hat{T}$, $\hat{W}$, $\hat{Y}$ that affect electroweak gauge boson propagation~\cite{Peskin:1991sw, Maksymyk:1993zm, Barbieri:2004qk}. The gauge self-energy function may be expanded in momentum as
\begin{equation}
\Pi_{VV'}(p^2) = \Pi_{VV'}(0) + p^2 \Pi_{VV'}'(0) + \frac{1}{2}p^4\Pi_{VV'}''(0) + \dots \,,
\end{equation}
with $\hat{T}$ depending only on combinations of $\Pi_{VV'}(0)$, while $\hat{S}$ also involves $\Pi_{VV'}'(0) $. Finally, the $\hat{W}$ and $\hat{Y}$ parameters are related to the $p^4$ term. In the SILH basis, these parameters have particularly simple representations in terms of the following dimension-6 operators~\cite{Barbieri:2004qk}
\begin{align}
\cO_{T} & = \frac{1}{2} ( H^{\dagger} \overleftrightarrow{D}_{\mu}H ) ( H^{\dagger} \overleftrightarrow{D}^{\mu}H ) \,,\label{eq:OT}  \\
\cO_{B+W} & = \frac{1}{4} \left[g_1 i( H^{\dagger} \overleftrightarrow{D}_{\mu} H ) \partial_{\nu} B^{\mu \nu} + g_2 i( H^{\dagger} \overleftrightarrow{D}_{\mu}^I H ) D_{\nu} W^{I\, \mu \nu} \right] \,, \label{BpW} \\
\cO_{2B} & = -\frac{1}{2}(\partial^{\mu}B_{\mu \nu}) (\partial_{\rho}B^{ \rho \nu}) \,, \label{eq:2B} \\
\cO_{2W} & = -\frac{1}{2}(D^{\mu}W_{\mu \nu}^I) (D_{\rho}W^{I \rho \nu}) \label{eq:2W}\,, 
\end{align}
where the relation to the oblique parameters is
\begin{align}
\hat{S} &= m_W^2 \cS_{B+W} (m_Z), & \hat{T} &= v^2 \cS_T (m_Z)\,, \\
\hat{W} &= m_W^2 \cS_{2W}(m_Z), & \hat{Y}  &= m_W^2 \cS_{2B}(m_Z)\,.
\end{align}
Note that instead of the usual $\cO_W$ and $\cO_{B}$ SILH operators we have defined the linear combinations $\cO_{B+W}$ and $\cO_{B-W}$ such that $\cS_{B} = (\cS_{B+W} + \cS_{B-W})/2$ and $\cS_{W} = (\cS_{B+W} - \cS_{B-W})/2$. This basis is chosen to be orthogonal at $m_Z$ such that $\hat{S}$ is given entirely by $\cS_{B+W}$, while $\cS_{B-W}$ does not contribute to the oblique parameters at leading order. However, the RGE does not preserve this orthogonality, so in general a contribution to $\cO_{B+W}$ from $\cO_{B-W}$ can arise at the loop level, allowing the $Z$-pole to also constrain heavy new physics generating $\cO_{B-W}$.

As is well known, the $\hat{S}$ and $\hat{T}$ parameters are strongly constrained by the pole observables. On the other hand, $\hat{W}$ and $\hat{Y}$ receive competitive constraints both on and off the pole, as shown recently~\cite{Greljo:2024ytg}. This is expected from our earlier accuracy complements energy discussion as being due to a competition between the energy enhancement these parameters receive above-pole versus the high statistics on pole. We have checked that it is an excellent approximation to neglect the running of these 4 operators between the above-pole energies down to $m_Z$. We therefore treat the Wilson coefficients entering the map to $\hat{S}, \hat{T}, \hat{W}, \hat{Y}$ as evaluated at $m_Z$ and do the combined on-and-above pole fit. We find that all the oblique parameters can be constrained at the $10^{-5}$ level
\begin{equation}
\left(
\begin{array}{cc}
\hat{S}  \\
\hat{T} \\
\hat{W} \\
\hat{Y} \\
\end{array}
\right)= \pm
\left(
\begin{array}{cc}
2.26 \\
1.16 \\
0.49 \\
1.90 \\
\end{array}
\right)\times 10^{-5}\,,
\label{eq:STWY}
\end{equation}
and their correlation matrix is
\begin{equation}
\rho = \left(
\begin{array}{cccc}
 1 & 0.795 & 0.331 & 0.868 \\
 0.795 & 1 & 0.088 & 0.461 \\
 0.331 & 0.088 & 1 & 0.256 \\
 0.868 & 0.461 & 0.256 & 1 \\
\end{array}
\right)\,.
\label{eq:STWYcorr}
\end{equation}
These results are in good agreement with Ref.~\cite{Greljo:2024ytg}. Having done the oblique parameter fit for completeness, in what follows we will drop the operators corresponding to $\hat{S}$ and $\hat{T}$ as any new physics model generating sizeable contributions to them will be dominated by on-pole phenomenology. In that case, one can simply use the likelihood given by \cref{eq:STWY,eq:STWYcorr}. However in general it is important to note that the oblique parameters only correctly capture on-pole physics for flavour-universal theories, and only at leading order. For non-universal theories or when the pole observables are computed at NLO, there are NP contributions that cannot be absorbed into the oblique parameters, and the SMEFT framework is required.

\subsubsection{Triple-gauge couplings}
\begin{figure}[t]
    \centering
    \hspace{-4mm}
    \includegraphics[scale=0.83]{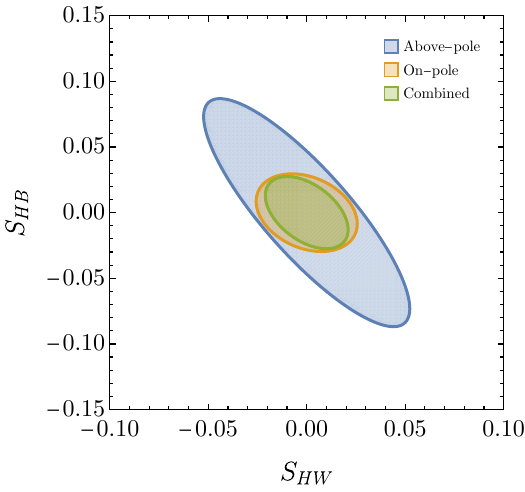} \hspace{3mm}
    \includegraphics[scale=0.83]{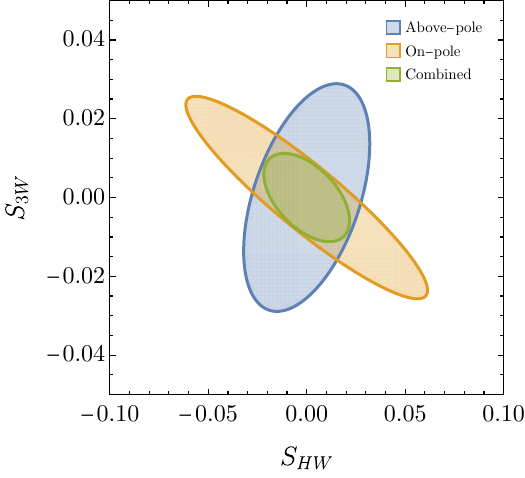}
    \caption{Projected FCC-ee sensitivities at 68\% CL to the three SMEFT operators contributing to anomalous triple-gauge couplings on and off the $Z/W$ pole.}
    \label{fig:TGCplot}
\end{figure}
Having examined corrections to gauge 2-point functions, we now move to 3-point functions similarly to the Higgs sector analysis. In the framework of anomalous couplings, there are 5 parameters that correct triple-gauge boson vertices. However, electroweak gauge invariance implies that only three are independent~\cite{Bobeth:2015zqa,Falkowski:2015jaa}. These anomalous triple-gauge couplings (aTGC) can receive contributions from three SMEFT operators in the SILH basis~\cite{Bobeth:2015zqa,Durieux:2017rsg},
\begin{align}
\cO_{HW} & = i(D^\mu H)^\dagger \tau^I (D^\nu H) W^I_{\mu\nu} \,, \label{eq:opHW}\\
\cO_{HB} & = i(D^\mu H)^\dagger (D^\nu H) B_{\mu\nu} \,, \\
\cO_{3W} &= \eps_{IJK} W_{\mu}^{I\nu} W_{\nu}^{J\,\rho} W_{\rho}^{K\,\mu} \label{eq:op3W}\,.
\end{align}
These operators may be directly probed at leading order in $e^+ e^- \rightarrow W^+ W^-$ processes at FCC-ee. In addition, using the SILH to Warsaw basis map in~\cite{Stefanek:2024kds}, we find
\begin{align}
\cO_{HB}  &= \cO_{B} - \frac12 \hyp_h g_1 Q_{H B} - \frac14 g_2  Q_{H WB} \,,\\
\cO_{HW} &= \cO_{W}  - \frac14 g_2 Q_{H W} - \frac12 \hyp_h g_1  Q_{H WB} \,,
\end{align}
where $\cO_{B,W}$ are SILH operators related to the $\hat{S}$ parameter. In fact, these mappings ensure that $\hat{S}$ receives an equal and opposite contribution from $\cO_{B,W}$ and $Q_{H WB}$ such that $\cO_{HB,HW}$ do not contribute to $Z$-pole observables at leading order. Note that the contribution to the Warsaw operators $Q_{HW}$ and $Q_{HB}$ remains, allowing sensitivity to $\cO_{HB,HW}$ in $e^+ e^- \rightarrow ZH$ processes. The fact that Higgs couplings can be used to break otherwise flat directions in the aTGC fit was discussed in Refs.~\cite{Pomarol:2013zra, Corbett:2013pja, Masso:2014xra, Ellis:2014dva, Ellis:2014jta, Dumont:2013wma, Falkowski:2015jaa} in the context of the LHC. Here we point out that the degeneracy can be broken by additional sensitivity to all three aTGC operators on the $Z$-pole at NLO, as depicted by the graph in~\cref{fig:aTGCdiagrams}. Some contributions can be seen from~\cref{eq:HWBrge}, which shows that all aTGC operators mix under RGE into $Q_{H WB}$ ($\hat{S}$ in the Warsaw basis). There are also finite contributions to $Z$-pole observables from all three operators in the NLO computation of Ref.~\cite{Bellafronte:2023amz} that we are using here.

In~\cref{fig:TGCplot} we show the projected FCC-ee sensitivities at 68\% CL to the three SMEFT operators affecting aTGC, on and off the pole as well as the combined sensitivity. We observe good complementarity, with the $Z$-pole performing slightly better for $\cO_{HB}$. On the other hand, the sensitivity is comparable on and off the pole for $\cO_{HW}$ and $\cO_{3W}$. Their anti-correlation in~\cref{fig:TGCplot} (right) further illustrates the power of combining $Z$-pole observables with the higher energy runs. Here it would also be interesting to include flavour data in the future, as aTGC also contribute to $b\rightarrow s\gamma$ and $b\rightarrow sl^+ l^-$ transitions at NLO~\cite{Bobeth:2015zqa}.

\subsubsection{Gauge sector combined fit}
\begin{figure}[t]
    \centering
    \hspace{-4mm}
    \includegraphics[scale=0.84]{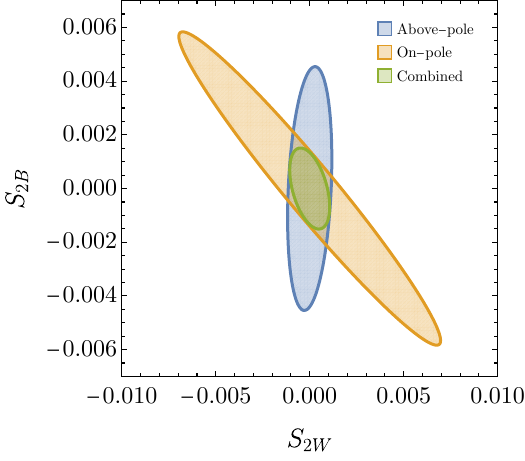}\hspace{3mm}
    \includegraphics[scale=0.82]{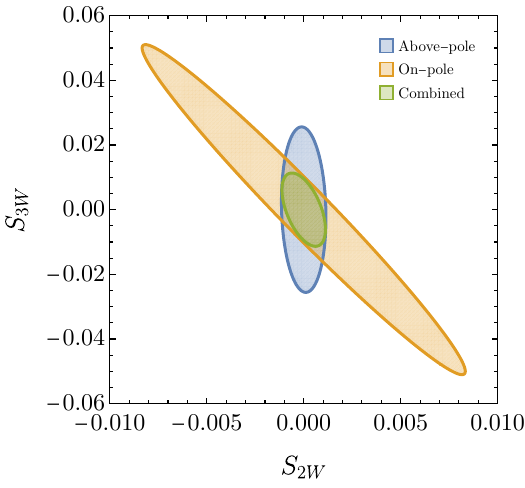}
    \caption{Projected FCC-ee sensitivities at 68\% CL to SMEFT operators correcting gauge 2- and 3-point functions on and off the $Z/W$ pole. Left: $S_{2W}$ and $S_{2B}$ corresponding to the $\hat W$ and $\hat Y$ parameters. Right: Correlation and constraints on the gauge 2- and 3-point functions $S_{2W}$ vs $S_{3W}$. }
    \label{fig:WY}
\end{figure}
The combined fit to all gauge sector operators in the SILH basis is reported here for convenience. We exclude $\mathcal{O}_{B+W}$ and $\mathcal{O}_T$ corresponding to the $\hat{S}$ and $\hat{T}$ parameters since these are completely dominated by the pole constraints. If one has a model that activates these operators at tree level, they will dominate the phenomenology and one should use the likelihood in~\cref{eq:STWY}. The rest of the gauge operators have comparable constraints on and off the pole (see~\cref{fig:WY}) and thus it is justified to do the combined fit to this subset. Taking $\Lambda = 1$ TeV as always, we find
\begin{equation}
\left(
\begin{array}{cc}
S_{2W} \\
S_{2B}  \\
S_{3W} \\
S_{HW} \\
S_{HB} \\
\end{array}
\right)= \pm
\left(
\begin{array}{cc}
0.08 \\
0.22 \\
1.52 \\
2.23 \\
2.34 \\
\end{array}
\right)\times 10^{-2}\,,
\end{equation}
with the following correlation matrix
\begin{equation}
\rho = \left(
\begin{array}{ccccc}
 1 & 0.10 & -0.34 & -0.07 & -0.06 \\
 0.10 & 1 & -0.74 & -0.42 & 0.16 \\
 -0.34 & -0.74 & 1 & -0.09 & 0.28 \\
 -0.07 & -0.42 & -0.09 & 1 & -0.69 \\
 -0.06 & 0.16 & 0.28 & -0.69 & 1 \\
\end{array}
\right)\,.
\end{equation}

We plot in \cref{fig:WY} the on-pole and above-pole projected 68\% CL sensitivities at FCC-ee in the plane of $S_{2W}$ vs. $S_{2B}$, corresponding to the $W$ and $Y$ parameters, on the left and for $S_{2W}$ vs $S_{3W}$ on the right. We see overall comparable constraints in the 2-dimensional ellipses, with off(on)-pole data giving stronger bounds along the $S_{2W}$ ($S_{2B}$) direction and on-pole data being crucial for restricting the overall combination to the green ellipse.

\subsection{4-fermion contact interactions}
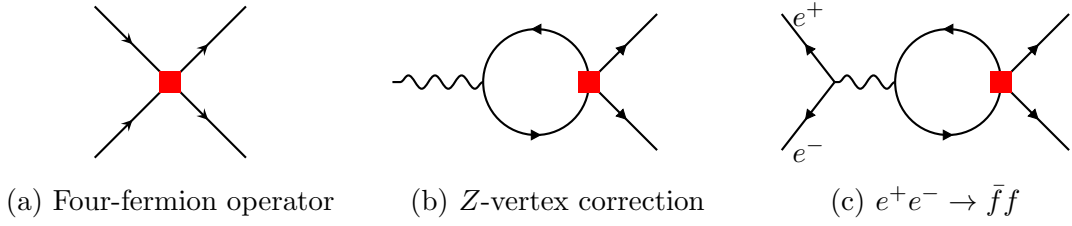
\begin{figure}[t!]
    \centering
    \begin{tabular}{c@{\hskip 1cm}c@{\hskip 1cm}c}
        \begin{tikzpicture}[thick,>=stealth,baseline=-7ex]
            \draw[fermion] (-1,-1) -- (0,0);
            \draw[fermion] (-1, 1) -- (0,0);
            \draw[fermion] (0, 0) -- (1,1);
            \draw[fermion] (0, 0) -- (1,-1);
            \node[wcsq] at (0,0) {};
             %%%%%%%%%%Labels
            % \node[left] at (-1.7,0) {$Z$};
            % \node[right] at (1.7,0) {$Z$};
            % \node[above] at (0,0.7) {$t$};
            % \node[below] at (0,-0.7) {$t$};
            % \node[right] at (-0.7,0.0) {$\delta g_{R33}^{Zu}$};
        \end{tikzpicture}
        & \hspace{-1cm}
        \begin{tikzpicture}[thick,>=stealth,baseline=-7ex]
            \draw[midarrow] (0.9,0) -- (1.8,0.9);
            \draw[midarrow] (0.9,0) -- (1.8,-0.9);
            \draw[midarrow] (0.9,0) arc (0:180:0.7);
            \draw[midarrow] (-0.5,0) arc (180:360:0.7);
            \draw[vector] (-0.5,0) -- (-1.7,0);
            \node[wcsq] at (0.9,0) {};
        \end{tikzpicture}
        &
        \begin{tikzpicture}[thick,>=stealth,baseline=-7ex]
            \draw[midarrow] (0.9,0) -- (1.8,0.9);
            \draw[midarrow] (0.9,0) -- (1.8,-0.9);
            \draw[midarrow] (0.9,0) arc (0:180:0.7);
            \draw[midarrow] (-0.5,0) arc (180:360:0.7);
            \draw[vector] (-0.5,0) -- (-1.3,0);
            \node[wcsq] at (0.9,0) {};
            \draw[midarrow] (-1.3,0) -- (-2,0.9) node[right] {$e^+$};
            \draw[midarrow] (-1.3,0) -- (-2,-0.9) node[right] {$e^-$};
        \end{tikzpicture}
        \\
        (a) Four-fermion operator & (b) $Z$-vertex correction & (c) $e^+ e^- \rightarrow \bar f f$
    \end{tabular}
    \caption{Four-fermion contact interaction that may contribute at LO to $e^+ e^- \rightarrow \bar f f$ (a), its contribution to $Z$ vertex corrections at NLO in (b), and its contribution to $e^+ e^- \rightarrow \bar f f$ at NLO (c). Red squares indicate SMEFT contributions to the vertex.}
    \label{fig:4Fdiagrams}
\end{figure}

The on- and above-pole sensitivity to four-fermion (4F) operators has already received significant attention in the literature. For example, the $Z$-pole sensitivity to 4F operators, especially those involving top quarks, has been studied in~\cite{Dawson:2019clf, Dawson:2022bxd,Allwicher:2023aql,Garosi:2023yxg,Allwicher:2023shc,Bellafronte:2023amz,Stefanek:2024kds,Haisch:2024wnw, Ge:2024pfn}.  On the other hand, above-pole sensitivity in the $e^+ e^- \rightarrow \bar f f$ processes at the $WW$, $ZH$, and $t\bar{t}$ energies was studied recently in Ref.~\cite{Greljo:2024ytg} via a dedicated flavor tagging analysis. Here we review and summarise the main conclusions. The projected sensitivities are shown in Fig.~\ref{fig:heatMap} for the four-quark (4Q), four-lepton (4L), and two-quark-two-lepton (2Q2L) operators, with darker shading indicating a stronger sensitivity to the effective scale of new physics $\Lambda^i_{\rm eff} = (\cC_i)^{-1/2}$, while~\cref{fig:4Fdiagrams} summarizes the relevant diagrams.

\subsubsection*{Semi-leptonic (2Q2L) operators}
For the semi-leptonic (2Q2L) operators, we see better sensitivity from on-pole data for operators involving top quarks and no electrons, namely: $[\mathcal{C}^{(1)}_{lq}]_{3333}$, $[\mathcal{C}^{(3)}_{lq}]_{3333}$, $[\mathcal{C}_{lu}]_{3333}$, $[\mathcal{C}_{eu}]_{3333}$, and $[\mathcal{C}_{qe}]_{3333}$. These operators run strongly into EW vertex corrections $\propto N_c y_t^2$ by closing the top loop and attaching two Higgses. On the other hand, above-pole data unsurprisingly performs better for operators involving electrons as they enter at LO and receive an energy enhancement. Still, a competitive bound can be achieved at the $Z$-pole if the operator has top quarks such as: $[\mathcal{C}^{(1)}_{lq}]_{1133}$, $[\mathcal{C}^{(3)}_{lq}]_{1133}$, $[\mathcal{C}_{lu}]_{1133}$, $[\mathcal{C}_{eu}]_{1133}$, and $[\mathcal{C}_{qe}]_{3311}$. Finally, other semi-leptonic operators without tops or electrons contribute to both on- and above-pole observables only at the loop level (see~\cref{fig:4Fdiagrams}), typically with comparable or slightly better sensitivity on the pole.

\subsubsection*{Four quark (4Q) operators}
Four quark operators enter via the loop diagrams in~\cref{fig:4Fdiagrams} for both on-pole and above-pole measurements, with on-pole data providing better sensitivity for operators involving top quarks such as $[\mathcal{C}^{(1)}_{qq}]_{3333}$, $[\mathcal{C}^{(1)}_{qu}]_{3333}$, and $[\mathcal{C}_{uu}]_{3333}$.\footnote{In the case of $[\mathcal{C}_{uu}]_{3333}$, the contribution to $Z$-pole observables occurs starting at the 2-loop level, but it is nonetheless phenomenologically relevant~\cite{Allwicher:2023aql,Allwicher:2023shc,Stefanek:2024kds}.} In the absence of tops, the dominant contribution comes from EW gauge running both on and off the pole. In this case, the constraints are typically comparable which can be understood by considering the energy enhancement above-pole vs. statistics on-pole, again demonstrating the principle that accuracy complements energy.

\subsubsection*{Four lepton (4L) operators}
Finally, four-lepton operators contribute at leading order to $e^+ e^- \rightarrow \bar f f$ if they involve electrons, in which case these observables dominate. A partial exception is $[\mathcal{C}_{ll}]_{1122}$ which also receives a noteable on-pole constraint due to its running into $[\mathcal{C}_{ll}]_{1221}$.  Otherwise, we are again in the situation of loop contributions via gauge running, leading to comparable constraints. In particular, this includes the pure third-generation case with operators such as $[\mathcal{C}_{ll}]_{3333}$, $[\mathcal{C}_{ee}]_{3333}$ and $[\mathcal{C}_{le}]_{3333}$.

%------------------------------------------------
\section{Specific UV model examples}
\label{sec:models}
%------------------------------------------------
Having illustrated the principle that accuracy complements energy in the SMEFT, we now move to examples in concrete UV-complete models. While we focus here on bosonic models, example models showing the power of combining on- and above-pole sensitivity to four fermion operators can be found in Ref.~\cite{Greljo:2024ytg}.

\subsection{Real singlet scalar}

We first consider the SM extended by a real singlet scalar $\phi$ with a $\mathds{Z}_2$ symmetry. The Lagrangian reads
\begin{equation}
\mathcal{L}_\phi = \frac{1}{2}\left(\partial_\mu\phi\right)^2 - \frac{1}{2}m_{\phi}^2\, \phi^2 -\frac{1}{2}\kappa |H|^2 \phi^2 - \frac{1}{4!} \lambda_\phi \phi^4 \, .
\end{equation}
Integrating out $\phi$ at 1 loop generates finite contributions to only two operators, namely $Q_{H\Box}$ and $Q_H$. The matching conditions at the scale $m_\phi$ are~\cite{Haisch:2020ahr,Fuentes-Martin:2022jrf}
\begin{align}
\cC_{H\Box} &= -\frac{1}{16\pi^2} \frac{\kappa^2}{24 m_\phi^2}, & \cC_{H} = -\frac{1}{16\pi^2} \frac{\kappa^3}{12 m_\phi^2}\,.
\end{align}
After electroweak symmetry breaking, these operators modify the Higgs kinetic term and self coupling, respectively. Thus, $\cC_{H\Box}$ induces a universal rescaling of all Higgs couplings after the kinetic term is canonically normalised. In addition, the physical mass of the scalar becomes $M_\phi^2 = m_S^2 + \frac{1}{2} \kappa v^2$, so it is convenient to define $f = \kappa v^2 / 2 M_\phi^2$ which measures how much of the total mass of the scalar comes from the cross quartic with the Higgs. If $f > 0.5$, we consider the field a ``loryon" as defined in~\cite{Banta:2021dek}. It is an interesting question whether the SM particles are the only ones that get most of their mass from the Higgs or if there exist BSM fields that share this property. Given the non-decoupling nature of these loryons with a finite parameter space, answering this fundamental question is a natural target for future colliders, and the singlet scalar is the last remaining open window that is difficult to probe completely~\cite{Crawford:2024nun}. 

\begin{figure}[t]
    \centering
    \includegraphics[scale=1]{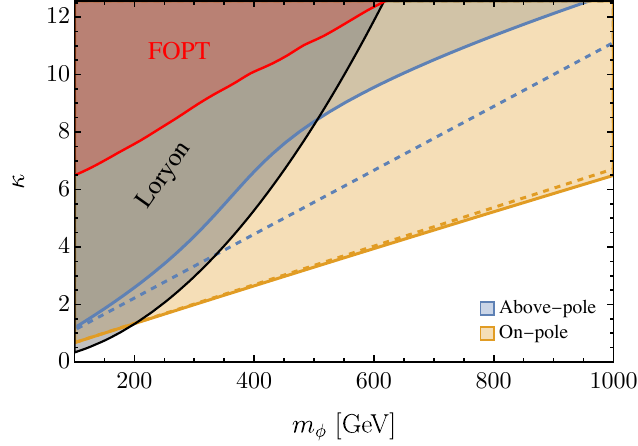}
    \caption{FCC-ee sensitivity at 68\% CL on the real singlet scalar model. The dashed lines show the limits if only $\cC_{H\Box}$ is matched, which neglects the correlated contribution from $\cC_H$. For parameters in the red region, the model can give rise to a first-order EW phase transition (curve taken from~\cite{Crawford:2024nun}).}
    \label{fig:RSSplot}
\end{figure}

In~\cref{fig:RSSplot}, we show the on- and above-pole constraints on the real singlet scalar model expected at FCC-ee, both of which exclude a first-order electroweak phase transition for the mass range shown. In~\cref{fig:RSSplot}, we show the constraints on the real singlet scalar model expected at FCC-ee, running from a fixed renormalization scale $\Lambda=\SI{500}{\giga\electronvolt}$ for simplicity. Both the on- and above-pole sets of observables would exclude a first-order electroweak phase transition for the mass range displayed. In particular, virtually all the loryon parameter space can now be probed by the inclusion of on-pole data that significantly extends the above-pole bounds. The dashed blue line shows the above-pole constraints assuming only $\cC_{H\Box}$ in the fit (for comparison to previous analyses), while the solid blue line is the full result also including $\cC_H$. The weaker constraint from the full result can be understood as follows. First, we have $\cC_H/\cC_{H\Box} = 2\kappa$, so the importance of $\cC_H$ relative to $\cC_{H\Box}$ grows linearly with $\kappa$. Next, these two Wilson coefficients contribute to the $ZH$ cross section with the ratio $\cC_H/\cC_{H\Box} \approx -0.07$. This means that for $\kappa \approx 7$, the NLO contribution of $\cC_H$ becomes of the same size as the LO contribution of $\cC_{H\Box}$ and there is a partial cancellation that leads to the weakened bound shown in~\cref{fig:RSSplot}.

On the other hand, the on-pole constraint given by the orange region in~\cref{fig:RSSplot} is dominated by the sizeable running of $Q_{H\Box} \rightarrow Q_{HD}$ given in~\cref{eq:CHboxRGE}, which in this model amounts to a 2-loop contribution to the $Z$-pole oblique parameters. We note that it is consistent to combine 1-loop matching with 1-loop running in this case due to the lack of any tree-level matching contributions. Performing a fit to all the on-pole data, we find a single operator 68\% CL bound of
\begin{equation}
\cC_{H\Box}( 1~\text{TeV}) \lesssim \frac{1}{(9.2~\text{TeV})^2} \,,
\label{eq:CHboxSingleOp}
\end{equation}
which yields $M / \kappa > 150$ GeV, in approximate agreement with~\cref{fig:RSSplot}.

We conclude this section by noting that the reliability of the dimension-6 SMEFT should not be trusted when $m_\phi \sim v$, as corrections from $d > 6$ operators become important. This signals the breakdown of the EFT and a computation in the full theory should be performed to correctly determine the small $m_\phi$ behavior. In particular, capturing the $Q_{H\Box} \rightarrow Q_{HD}$ contribution to the oblique parameters will require a 2-loop calculation in the full theory, which we plan to perform in future work. This will be necessary to determine if the low mass regime of the loryon parameter space can be fully excluded.

%%%%%%%%%%%%%%%%%%%%%%%%%%%%%%%%%%%%%%%%%%%%%%%%%
\subsection{Weakly interacting massive particles}%
\begin{figure}[t]
    \centering
    \includegraphics[scale=0.8]{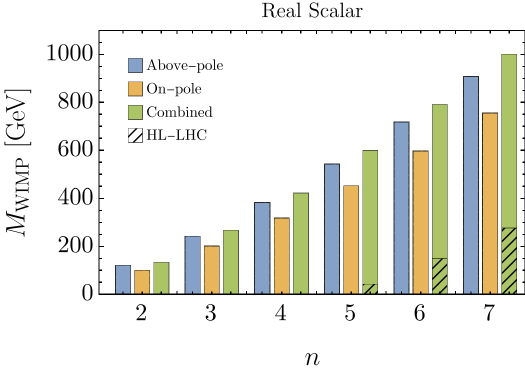} \hspace{2.5mm}
    \includegraphics[scale=0.82]{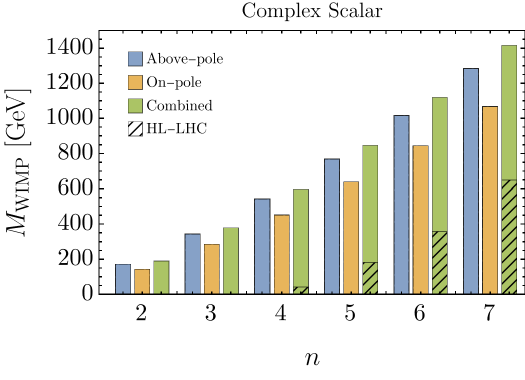}
    \caption{FCC-ee 95\% CL exclusion for scalar WIMPs without hypercharge.}
    \label{fig:WIMPscalar}
\end{figure}
\begin{figure}[t]
    \centering
    \hspace{-3.5mm}
    \includegraphics[scale=0.8275]{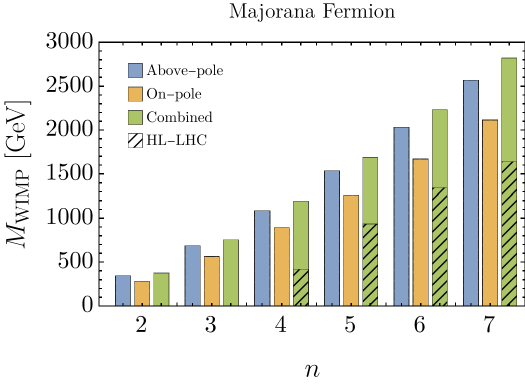} \hspace{2.5mm}
    \includegraphics[scale=0.8275]{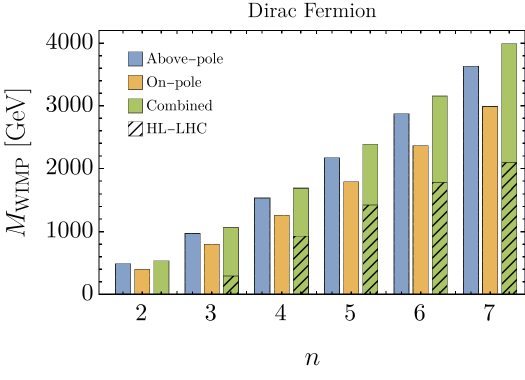}
    \caption{FCC-ee 95\% CL exclusion for fermionic WIMPs without hypercharge.}
    \label{fig:WIMPfermion}
\end{figure}

Electroweak multiplets are motivated in many BSM extensions but are difficult to search for at hadron colliders, with current indirect exclusions in Drell-Yan being insensitive to low $n$ multiplets and probing only the higher multiplets in the $\mathcal{O}(100)$ GeV range up to around 2 TeV for a 7-plet Dirac fermion~\cite{DiLuzio:2018jwd}. Heavy Weakly Interacting Massive Particles (WIMPs) therefore form a current blind spot in BSM searches that could be indirectly explored at FCC-ee~\cite{Cirelli:2005uq,Harigaya:2015yaa,DiLuzio:2018jwd,Bottaro:2021snn, Bottaro:2022one}.  

The matching conditions for an $SU(2)_L$ $n$-tuplet WIMP with hypercharge $Y$ to the SILH basis operators as normalised in~\cref{eq:2B,eq:2W,eq:op3W} are
\begin{align}
\cS_{2B} &= \frac{g_1^2}{16\pi^2} \frac{n Y^2}{30M_{\rm WIMP}^2} N_{2} , & \cS_{2W} &= \frac{g_2^2}{16\pi^2} \frac{n(n^2-1)}{360M_{\rm WIMP}^2} N_{2}, & \cS_{3W} = \frac{g_2^3}{16\pi^2} \frac{n(n^2-1)}{2160 M_{\rm WIMP}^2} N_{3} \,,
\label{eq:WIMPmatching}
\end{align}
where the counting factors are $N_{2} = 1/2,1,4,8$ and $N_{3} = 1/2,1,-1,-2$ for real scalars, complex scalars, Majorana fermions, and Dirac fermions, respectively. As discussed in the above EFT analysis, these operators can be constrained by both above-pole and on-pole data where we expect similar sensitivity and an improved combined bound. This is shown in Figs.~\ref{fig:WIMPscalar} and ~\ref{fig:WIMPfermion} for real and complex scalar WIMPs and Majorana and Dirac fermion WIMPs, with zero hypercharge. In all cases, we run down from $\Lambda = \SI{1}{\tera\electronvolt}$ to the IR scale of the corresponding observable. The blue and orange bars denote the projected 95 \% CL sensitivity of above-pole and on-pole data to the mass of the WIMP in GeV for various $n$-plets ranging from $n=2$ to 7. The green bar is the combined sensitivity reach, with the black hatched bars indicating the HL-LHC bounds from the study in~\cite{DiLuzio:2018jwd}. We see that in all cases FCC-ee can significantly extend the HL-LHC exclusions.

\subsection{Custodial weak quadruplet}
\begin{figure}[t]
    \centering
    \hspace{-4.5mm}
    \includegraphics[scale=0.8]{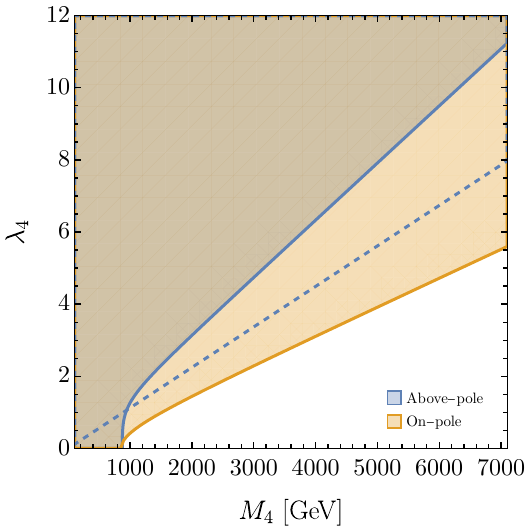}\hspace{4.5mm}
    \includegraphics[scale=0.826]{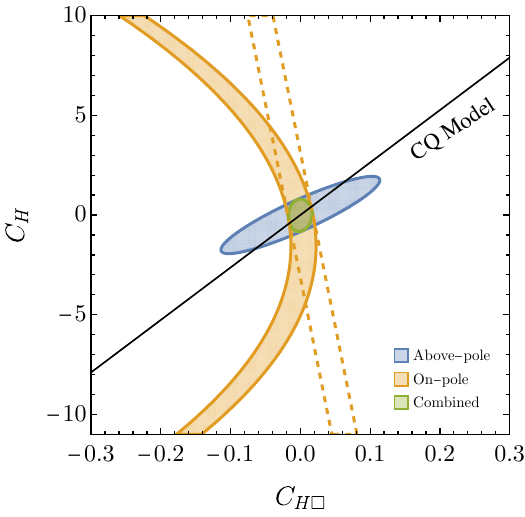} 
    \caption{Left: FCC-ee sensitivity at 68\% CL to the custodial weak quadruplet model. For comparison, the dashed blue line shows the above-pole sensitivity considering only the tree-level matching to $|H|^6$. Right: The black line in the $C_{H\Box}$ vs $C_H$ plane corresponds to the CQ model. }
    \label{fig:CQplot}
\end{figure}
Finally, as an example of the interplay between the Higgs and gauge sectors, we consider the custodial quadruplet (CQ) model as defined in~\cite{Durieux:2022hbu}. This model is motivated by allowing for a naturally larger ratio of Higgs self-coupling to Higgs coupling modifications, thus in principle showing effects first in the former rather than the latter. However, we will see that both modifications must be included despite one being loop suppressed in order to obtain accurate projected sensitivities. 

Following the notation of~\cite{deBlas:2017xtg}, this model contains the complex scalars $\Theta_1 \sim ({\bf 1},{\bf 4})_{1/2}$ and $\Theta_3 \sim ({\bf 1},{\bf 4})_{3/2}$. These states, in addition to being WIMPs, may couple linearly to three SM Higgses, generating only $|H|^6$ when integrated out at tree level. However, in general these Higgs couplings violate custodial symmetry and thus the operator $Q_{HD} = |H^{\dagger}D_\mu H|^2$ is generated as a finite 1-loop matching contribution, dominating the phenomenology. It can be forbidden by embedding $\Theta_{1,3}$ into a custodial multiplet, specifically a $({\bf 4},{\bf 4})$ of $SU(2)_L \times SU(2)_R$. Custodial symmetry then enforces these states to have the same mass $M_4$ and coupling $\lambda_4$ to the Higgs, canceling the contribution to $Q_{HD}$. The relevant custodially symmetric UV Lagrangian is
\begin{equation}
\mathcal{L}_{\rm CQ} \supset -M_4^2 \left( |\Theta_1|^2 + |\Theta_3|^2\right) -\lambda_4 \left(H^*H^*(\varepsilon H) \Theta_1 + \frac{1}{\sqrt{3}}H^*H^*H^* \Theta_3 \right) +{\rm h.c.} \,.
\end{equation}
With the aid of the~\texttt{Matchete} software~\cite{Fuentes-Martin:2022jrf}, the full 1-loop matching to the dimension-6 SMEFT at the scale $M_4$ for custodial quadruplet model can be written in two lines
\begin{align}
&\mathcal{L}_{\rm{CQ}}^{\rm d=6} = \frac{2}{3}\frac{\lambda_4^2}{M_4^2}\left(1 + \frac{21\lambda_{\rm SM}}{16\pi^2}  \right)|H|^6 - \frac{\lambda_4^2}{4\pi^2 M_4^2} \partial_\mu |H|^2 \partial^\mu |H|^2 -\frac{\lambda_4^2}{3\pi^2 M_4^2}|H|^2 (H^\dagger D^2 H + {\rm h.c.}) \nonumber \\
&+ \frac{1}{48\pi^2 M_4^2} \left[\frac{g_2^3}{3!}\eps_{IJK} W_{\mu}^{I\nu} W_{\nu}^{J\,\rho} W_{\rho}^{K\,\mu} - \frac{g_1^2}{2}(\partial^{\mu}B_{\mu \nu}) (\partial_{\rho}B^{ \rho \nu}) - \frac{g_2^2}{2} (D^{\mu}W_{\mu \nu}^I) (D_{\rho}W^{I \rho \nu}) \right]  \,,
\end{align}
where $\lambda_{\rm SM}$ is the SM Higgs quartic. The operator $\partial_\mu |H|^2 \partial^\mu |H|^2$ is equivalent to $-Q_{H\Box}$ via integration by parts. Furthermore, the terms containing $D^2 H$ can be reduced using the Higgs equation of motion to another $|H|^6$ term with coefficient $2\lambda_{\rm SM} \lambda_4^2 / (3\pi^2 M_4^2)$ plus dimension-6 Yukawa corrections of the form $\sum_{\psi} Y_{\rm \psi}^{ij} |H|^2 \bar\psi_L^i H \psi_R^j$. These terms correct Higgs decays, but we drop them in what follows since they are always suppressed by small SM Yukawa couplings $Y_{\rm \psi}$. Similarly, the total loop correction to $|H|^6$ proportional to $\lambda_{\rm SM}$ is of size $10^{-2}$ relative to the tree contribution and can be safely neglected. Thus for phenomenological purposes we may work in the limit where we keep matching contributions only in $\lambda_4, g_1$, and $g_2$. We note that the $\lambda_4$ terms of our result agree with those of~\cite{Durieux:2022hbu}, while the contributions to $\mathcal{O}_{2B,2W,3W}$ can be easily derived from~\cref{eq:WIMPmatching}.

In~\cref{fig:CQplot} (left), we show the sensitivity of FCC-ee to the custodial weak quadruplet model on and off the $Z/W$ pole, where we run from a fixed high-scale $\Lambda = \SI{1}{\tera\electronvolt}$ for simplicity. We compare this with the above-pole sensitivity obtained only by considering the tree-level modification of the Higgs self-coupling (dashed blue line). We see that the full above-pole result shows a significant reduction in sensitivity. This is because while the contribution to $\cC_{H\Box}$ is loop suppressed and $\cC_{H}$ is generated at tree level, $\cC_{H\Box}$ enters $\sigma(ZH)$ at LO while $\cC_{H}$ enters at NLO, with a relative minus sign. The CQ model predicts a fixed ratio of $\cC_{H}/\cC_{H\Box} = +8\pi^2 / 3$, so the two contributions partially cancel in the $ZH$ cross section. This can be seen in~\cref{fig:CQplot} (right), where the CQ model accidentally approximately aligns with the flattest above-pole direction. On the other hand, the $Z$-pole provides a strong bound in this direction, dominantly via the NLO contribution of $\cC_{H\Box}$ to $\cC_{HD}$. Finally, the $\lambda_4$-independent lower bound of $M_4 \gtrsim 800$ GeV comes from the $\hat{W}$ and $\hat{Y}$ parameters, which scale as $g_{\rm EW}^2/M_{4}^2$.

A contribution to $\cC_{HD}$ arises despite the fact that the NP coupling $\lambda_4$ respects custodial symmetry, because the same is not true of SM couplings such as $g_1$. At the 2-loop level we thus expect a contribution to $Q_{HD}$. Since $Q_H$ does not renormalise any bosonic operators apart from itself at the 2-loop level~\cite{Born:2024mgz}, it is consistent to combine our 1-loop matching computation with 1-loop running. We can therefore estimate the 2-loop contribution to $Q_{HD}$ via the first leading-log RG mixing of $Q_{H\Box} \rightarrow Q_{HD}$. From~\cite{Alonso:2013hga}, we get
\begin{equation}
\cC_{HD}(m_Z) = \frac{5 g_1^2}{24\pi^2} \cC_{H\Box}(M_4) \log\left( \frac{m_Z^2}{M_4^2}\right) = \frac{40}{3} \frac{g_1^2 \lambda_4^2}{(16\pi^2)^2 M_4^2} \log\left( \frac{m_Z^2}{M_4^2}\right)\,.
\label{eq:CHBox2CHD}
\end{equation}
For $\lambda_4 = 1$ and $M_{4} = 1$ TeV, this yields a contribution to the $T$ parameter of $\hat T = -v^2 \cC_{HD}(m_Z)/2 \approx 10^{-5}$, which is at the detectable level according to~\cref{eq:STWY}. However, $Q_{HD}$ is not the only operator activated (see \emph{e.g.}~\cref{eq:CHl3RGE}), and using~\cref{eq:CHboxSingleOp} we find $M_4 / \lambda_4 \gtrsim 1.5$ TeV, in agreement with the on-pole limit in~\cref{fig:CQplot}.

%------------------------------------------------
\section{Discussion and conclusions} 
\label{sec:conclusions}
%------------------------------------------------

The importance of a Tera-$Z$ factory to a future programme of particle physics cannot be overstated. Far from being a mere repeat of LEP, the new regime of ultra-high precision accessible by circular $e^+ e^-$ colliders such as FCC-ee or CEPC enters the threshold to unlock a wider range of phenomena across all sectors of the SM through higher-order quantum effects. In a similar way that a hadron collider benefits from multiple initial state partons for probing different types of physics, multi-loop processes at the $Z$ pole democratically mixes in all accessible particles via quantum fluctuations.  This makes Tera-$Z$ uniquely suited to a general exploration of fundamental physics at the zepto-scale, with indirect sensitivity to a wide variety of generic BSM physics up to the tens of TeV in energy range. 

We have emphasised in this work how high-precision accuracy at Tera-$Z$ complements and can even beat high-energy measurements. When they are sensitive to the same new physics modifications entering at different orders or with energy-growing effects, the increased statistics compensate for higher loop suppression on the $Z$ pole or for energy enhancements above-pole. It is a simple counting exercise to see that $Z$-pole sensitivity at NLO is of the same order of magnitude as above-pole sensitivity at LO. Tera-$Z$ is therefore expected to provide competitive NLO sensitivity to above-pole measurements at LO for any Wilson coefficient that they both have access to. This ``accuracy complements energy" principle also applies at NNLO on-pole vs. NLO above-pole, as well as if a Wilson coefficient enters at LO both on-pole and above-pole but is energy-enhanced in the latter.  

We studied new physics contributions to the Higgs 2-point and 3-point functions, which modify the Higgs self-energy, self-coupling, and Higgs-vector-vector interactions, and to the gauge boson 2-point and 3-point functions which enter in the oblique parameters and anomalous triple-gauge boson couplings. The Higgs self-energy and vector-vector interactions contribute at leading order to $e^+ e^- \to ZH$, while the Higgs self-coupling enters this process at NLO. On the other hand, they all affect the $Z$ pole oblique parameters at NLO and NNLO respectively, leading as expected to competitive on-pole projected sensitivities at FCC-ee and complementary bounds when combined with above-pole data. We find a similar story for the anomalous triple-gauge couplings that enter at leading order in $WW$ and at NLO at the $Z$ pole. Finally, the $Z$ pole oblique parameters $\hat{W}$ and $\hat{Y}$ can be constrained at leading order both on- and above-pole, with the former being statistics-enhanced and the latter energy-enhanced, such that they also lead to an overall similar and complementary projected bounds.  

Since so many Wilson coefficients enter the oblique parameters at higher orders, global SMEFT fits beyond leading order will find large correlations between seemingly unrelated above-pole Wilson coefficients, and cancellations between them when marginalising will give the appearance of weaker bounds. Hence it is crucial to emphasise that with Tera-$Z$ it will be especially important to match to explicit UV models to extract useful information, as the dynamical information from the model will correlate operators and pick out specific directions in the SMEFT Wilson coefficient space that typically result in stronger bounds instead. The fundamental reason is because the volume of the region that can be constrained in Wilson coefficient space is reduced by combining datasets beyond leading order. We find that this is indeed the case for the $Z_2$-symmetric real singlet scalar, scalar and fermion WIMPs, and a custodial weak quadruplet scalar. These explicit UV models all benefit significantly from the inclusion of on-pole data expanding coverage of their projected parameter space at FCC-ee. 

To conclude, the principle of accuracy complements energy expands the physics case for circular $e^+ e^-$ machines featuring a Tera-$Z$ run as part of a comprehensive particle physics programme that not only includes studying the Higgs boson more precisely, but also involves exploring inclusively the landscape of generic new physics up to the tens of TeV range. This adds further weight to the conclusion that a far broader scientific scope than alternative options is possible at FCC-ee~\cite{Blondel:2024mry}. Unlocking a new quantum regime of ultra-high precision will come with new challenges that must be met by the experimental and theoretical communities; but if history teaches us anything it is that the value of extreme refinement in the fundamental sciences goes beyond adding a few decimal places to measurements.

%------------------------------------------------
\section*{Acknowledgments}
%------------------------------------------------
We thank Matthew McCullough, Pier Paolo Giardino, Admir Greljo, Alessandro Valenti, Felix Wilsch for useful dicussions. We thank Anders Eller Thomsen for his knowledge of group magic. The work of BAS is supported by a CDEIGENT grant from the Generalitat Valenciana with grant no. CIDEIG/2023/35. VM is supported by a KCL NMES faculty studentship. TY is supported by United Kingdom Science and Technologies
Facilities Council (STFC) grant ST/X000753/1.

\appendix
\section{SMEFT theory expression for $\sigma(e^+ e^- \rightarrow W^+ W^-$)}
\label{app:WW}

In this appendix, we give the results of our calculation of $\sigma(e^+ e^- \rightarrow W^+ W^-)$ at tree level for the relevant center of mass energies for FCC-ee, $E_{\rm{CM}} = \SI{240}{\giga \electronvolt}$ and $\SI{365}{\giga \electronvolt}$. We calculate this cross section using the Feynman Rules implemented in {\tt SMEFT-FR}~\cite{Dedes:2017zog,Dedes:2019uzs,Dedes:2023zws}, from which we generate a {\tt FeynArts}~\cite{Hahn:2000kx} model, that can then be imported into {\tt FeynCalc}~\cite{Mertig:1990an,Shtabovenko:2016sxi,Shtabovenko:2020gxv,Shtabovenko:2023idz} to calculate the final cross section, using the $\{M_W,M_Z,G_F\}$-scheme. We present our results first in the non-redundant Warsaw Basis defined in the WCXF-format~\cite{Aebischer:2017ugx} and make no flavor assumptions on the coefficients. With these choices, only 8 operators modify the cross section:
\begin{equation}
    \mathcal{C} = \{\mathcal{C}_W,\, \mathcal{C}_{HWB},\, \mathcal{C}_{HD},\,[\mathcal{C}_{He}]_{11},\, [\mathcal{C}_{Hl}^{(1)}]_{11},\, [\mathcal{C}_{Hl}^{(3)}]_{11},\, [\mathcal{C}_{Hl}^{(3)}]_{22},\, [\mathcal{C}_{ll}]_{1221}\}.
    \label{eq:ee_WW_LO_coeffs}
\end{equation}
 and their contribution to the total cross section can be parametrised as follows:
 \begin{equation}
 \frac{\Delta \sigma}{\sigma_{\rm{SM}}} = \sum_{i} b_i \mathcal{C}_i = \sum_{i} b_i\,\frac{C_i}{\Lambda^2} 
 \label{eq:WW_Relative_Contribution}
 \end{equation}
The numerical values of $b_i$ for both energies assuming $\Lambda = 1\TeV$, as well as the Standard Model value are given in Table \ref{tab:ee_WW}. 

\begin{table}[t]
    \centering
    \resizebox{\textwidth}{!}{%
    \begin{tabular}{|c|c|c|c|c|c|c|c|c|c|}\hline
       $E_{\rm{CM}}$ [$\si{\text{GeV}}$] & SM [$\si{\pico\barn}$] & $\mathcal{C}_W$& $\mathcal{C}_{HWB}$& $\cC_{HD}$&$[\cC_{He}]_{11}$& $[\cC_{Hl}^{(1)}]_{11}$& $[\cC_{Hl}^{(3)}]_{11}$& $[\cC_{Hl}^{(3)}]_{22}$& $[\cC_{ll}]_{1221}$\\\hline\hline
       240 & 17.1544& -6.18& 4.81& -5.00& -17.48& -27.07& 142.6& -121.2& 121.2\\\hline
       365  & 10.7387& -10.08& 8.68& -3.58& -24.20& -49.54& 166.7& -121.2& 121.2\\\hline
    \end{tabular}
    }
    \caption{Contribution $b_i \cdot 10^{3}$ of Wilson Coefficients $C_i/(\SI{1}{\tera\electronvolt})^2$ to  $\sigma(e^+ e^- \rightarrow W^+ W^-)$ normalised to the Standard Model value given in the second row for $E_{CM} = \SI{240}{\text{GeV}}$ and $\SI{365}{\text{GeV}}$, corresponding to the $ZH$- and $t\bar{t}$-thresholds.}
    \label{tab:ee_WW}
\end{table}

As described in Sec. \ref{sec:PureGauge}, it is sometimes more convenient to work in the SILH basis. Both share a large number of coefficients, but motivated by Higgs and gauge coupling modifications, as well as the situation described in \cite{Stefanek:2024kds}, we focus on the subset of coefficients introduced in Eqs. \ref{eq:OT}-\ref{eq:2W} and \ref{eq:opHW} -\ref{eq:op3W} and App. \ref{app:SILH_bounds}. With these assumptions, the only operators affecting the diboson production cross section:
\begin{equation}
    \mathcal{S} = \{\mathcal{S}_{3W},\, \mathcal{S}_{HB},\, \mathcal{S}_{HW},\,\mathcal{S}_{2B},\,\mathcal{S}_{2W},\,\mathcal{S}_{B+W},\,\mathcal{S}_{B-W},\,\mathcal{S}_{T}\}.
    \label{eq:ee_WW_LO_coeffs_SILH}
\end{equation}
Their relative contribution assuming $\Lambda = \SI{1}{\tera\electronvolt}$ is given in Table \ref{tab:ee_WW_SILH}. 

\begin{table}[t]
    \centering
    \resizebox{\textwidth}{!}{%
    \begin{tabular}{|c|c|c|c|c|c|c|c|c|c|}\hline
       $E_{\rm{CM}}$ [$\si{\text{GeV}}$] & SM [$\si{\pico\barn}$] & $\mathcal{S}_{3W}$ & $ \mathcal{S}_{HB}$ & $ \mathcal{S}_{HW}$ & $\mathcal{S}_{2B}$ & $\mathcal{S}_{2W}$ & $\mathcal{S}_{B+W}$& $\mathcal{S}_{B-W}$& $\mathcal{S}_{T}$\\\hline\hline
       240 & 17.1544& -6.18& 2.98& 3.03& -1.67& -14.99& 1.79& -0.45& 9.99\\\hline
       365  & 10.7387& -10.08& 6.08& 6.59& -2.91& -17.52& 3.73& -1.05 & 7.15\\\hline
    \end{tabular}
    }
    \caption{Same as Table \ref{tab:ee_WW}, but with our chosen subset of SILH basis operators.}
    \label{tab:ee_WW_SILH}
\end{table}
    
\section{Single operator sensitivities}
\label{app:B}
We report here the projected sensitivity of FCC-ee to SMEFT Wilson coeffficients (WCs) renormalised at $\Lambda = \SI{1}{\tera\electronvolt}$ when activating one WC at a time using the likelihood described in \ref{sec:setupobs}. Since we are setting the projected experimental values to their SM value and our theory predictions are linear in $\cC_i$, our likelihood is Gaussian and centered at $\cC_i = 0$. Thus, we quote the $2\sigma$ intervals in terms of $\Lambda_{\rm{eff}}^{\rm{i}} = \left|\left(\sqrt{\cC_i}\right)^{-1}\right|$ using observables measured on the $Z$ and $W$ pole (on-pole), as well as above-pole observables measured at the $ZH$ and $t\bar{t}$ thresholds (above-pole) and, lastly, their combined sensitivity.  For the on- and above-pole observables, we quote also the observables with the largest pull 
\begin{equation}
    p(O_{\rm{i}})_{\rm{j}} = \frac{O_i(\cC_j)-O_i(0)}{\sigma_i}
    \label{eq:pull}
\end{equation}
with $\cC_j$ evaluated at the $2\sigma$ limit. This can be interpreted as the observable that most strongly constrains the coefficient. For the combined sensitivity, since on- and above-pole are not correlated, the strongest bound gives the largest pull as well.
\subsection{Warsaw Basis}
\small
\begin{longtable}{|c|c|c|c|c|c|}
\caption{Single Operator $2\sigma$ confidence intervals in the Warsaw basis exceding \SI{1}{\tera\electronvolt} from On- and Off-Pole observables, as well as their combined projected sensitivity. The Wilson Coefficients are renormalized at $\Lambda=\SI{1}{\tera\electronvolt}$. We also report the most constraining measurement for each coefficient for On- and Off-Pole Observables.} \\
\hline
\toprule
$\mathcal{C}_{\rm{i}}$ & $\Lambda^{\rm{eff}}_{\rm{i}}$ on-pole & Obs on-pole & $\Lambda^{\rm{eff}}_{\rm{i}}$ off-pole & Obs off-pole & $\Lambda^{\rm{eff}}_{\rm{i}}$ Combined \\
\hline
\midrule
\endfirsthead
\caption[]{$2\sigma$ CL in the Warsaw basis. Coefficients are renormalized at $\Lambda=\SI{1}{\tera\electronvolt}$} \\
\hline
\toprule
$\mathcal{C}_{\rm{i}}$ & $\Lambda^{\rm{eff}}_{\rm{i}}$ on-pole & Obs on-pole & $\Lambda^{\rm{eff}}_{\rm{i}}$ off-pole & Obs off-pole & $\Lambda^{\rm{eff}}_{\rm{i}}$ Combined \\
\hline
\midrule
\endhead
\midrule
\multicolumn{6}{r}{Continued on next page} \\
\hline
\midrule
\endfoot
\bottomrule
\endlastfoot
$\mathcal{C}_{HWB}$ & 111.55 & $A^{Z}_e$ & 7.01 & $\sigma_{Zh}^{240}$ & 111.55 \\
\hline
$[\mathcal{C}_{Hl}^{(3)}]_{22}$ & 67.88 & $A^{Z}_e$ & 26.34 & $\sigma_{WW}^{240}$ & 68.25 \\
\hline
$[\mathcal{C}_{Hl}^{(3)}]_{11}$ & 65.11 & $A^{Z}_\mu$ & 28.34 & $\sigma_{WW}^{240}$ & 65.69 \\
\hline
$\mathcal{C}_{HD}$ & 57.11 & $m_W$ & 4.47 & $\sigma_{WW}^{240}$ & 57.11 \\
\hline
$[\mathcal{C}_{ll}]_{1221}$ & 52.98 & $A^{Z}_e$ & 37.78 & $R_\mu^{240}$ & 56.13 \\
\hline
$[\mathcal{C}_{He}]_{11}$ & 54.23 & $A^{Z}_e$ & 13.48 & $\sigma_{Zh}^{365}$ & 54.28 \\
\hline
$[\mathcal{C}_{He}]_{22}$ & 53.58 & $A^{Z}_\mu$ & 1.41 & $R_\mu^{163}$ & 53.58 \\
\hline
$[\mathcal{C}_{Hl}^{(1)}]_{22}$ & 51.18 & $A^{Z}_\mu$ & 0.94 & $R_\mu^{163}$ & 51.18 \\
\hline
$[\mathcal{C}_{Hl}^{(1)}]_{11}$ & 50.85 & $A^{Z}_e$ & 14.97 & $\sigma_{Zh}^{365}$ & 50.94 \\
\hline
$[\mathcal{C}_{lq}^{(1)}]_{1133}$ & 12.52 & $A^{Z}_e$ & 48.63 & $R_b^{240}$ & 48.68 \\
\hline
$[\mathcal{C}_{lq}^{(3)}]_{1133}$ & 17.12 & $A^{Z}_\mu$ & 47.35 & $R_b^{240}$ & 47.55 \\
\hline
$[\mathcal{C}_{lq}^{(3)}]_{1111}$ & 6.84 & $A^{Z}_e$ & 47.49 & $R_c^{240}$ & 47.50 \\
\hline
$[\mathcal{C}_{lq}^{(1)}]_{1122}$ & 2.12 & $A^{Z}_e$ & 46.94 & $R_c^{240}$ & 46.94 \\
\hline
$[\mathcal{C}_{lq}^{(3)}]_{1122}$ & 6.81 & $A^{Z}_e$ & 45.57 & $R_c^{240}$ & 45.57 \\
\hline
$[\mathcal{C}_{He}]_{33}$ & 37.27 & $A^{Z}_\tau$ & 1.41 & $R_\tau^{163}$ & 37.27 \\
\hline
$[\mathcal{C}_{Hl}^{(3)}]_{33}$ & 36.59 & $A^{Z}_\tau$ & 2.21 & $R_\tau^{163}$ & 36.59 \\
\hline
$[\mathcal{C}_{eu}]_{1122}$ & 3.26 & $A^{Z}_e$ & 36.36 & $R_c^{240}$ & 36.36 \\
\hline
$[\mathcal{C}_{Hl}^{(1)}]_{33}$ & 35.72 & $A^{Z}_\tau$ & 0.94 & $R_\tau^{163}$ & 35.72 \\
\hline
$[\mathcal{C}_{ll}]_{1122}$ & 11.94 & $A^{Z}_e$ & 35.29 & $R_\mu^{240}$ & 35.41 \\
\hline
$[\mathcal{C}_{ll}]_{1133}$ & 2.30 & $A^{Z}_e$ & 35.29 & $R_\tau^{240}$ & 35.29 \\
\hline
$[\mathcal{C}_{ll}]_{1331}$ & 2.25 & $A^{Z}_e$ & 35.19 & $R_\tau^{240}$ & 35.19 \\
\hline
$[\mathcal{C}_{ee}]_{1122}$ & 2.67 & $A^{Z}_e$ & 33.10 & $R_\mu^{240}$ & 33.10 \\
\hline
$[\mathcal{C}_{ee}]_{1133}$ & 2.38 & $A^{Z}_e$ & 33.10 & $R_\tau^{240}$ & 33.10 \\
\hline
$[\mathcal{C}_{Hq}^{(3)}]_{11}$ & 31.01 & $\Gamma_Z$ & 1.90 & $\sigma_{Zh}^{240}$ & 31.01 \\
\hline
$[\mathcal{C}_{Hq}^{(1)}]_{33}$ & 30.78 & $m_W$ & 1.87 & $\sigma_{WW}^{240}$ & 30.78 \\
\hline
$[\mathcal{C}_{ed}]_{1133}$ & 2.27 & $A^{Z}_e$ & 30.65 & $R_b^{240}$ & 30.65 \\
\hline
$[\mathcal{C}_{eu}]_{1111}$ & 3.26 & $A^{Z}_e$ & 29.60 & $R_e^{365}$ & 29.61 \\
\hline
$[\mathcal{C}_{qe}]_{3311}$ & 13.20 & $A^{Z}_e$ & 26.50 & $R_b^{240}$ & 26.90 \\
\hline
$[\mathcal{C}_{Hu}]_{33}$ & 26.04 & $m_W$ & 1.94 & $\sigma_{WW}^{240}$ & 26.04 \\
\hline
$[\mathcal{C}_{Hq}^{(3)}]_{22}$ & 25.60 & $R^{Z}_\mu$ & 1.86 & $\sigma_{Zh}^{240}$ & 25.60 \\
\hline
$[\mathcal{C}_{Hq}^{(3)}]_{33}$ & 23.70 & $R^{Z}_\mu$ & 1.70 & $R_b^{240}$ & 23.70 \\
\hline
$[\mathcal{C}_{lu}]_{1122}$ & 3.04 & $A^{Z}_e$ & 23.66 & $R_c^{240}$ & 23.66 \\
\hline
$[\mathcal{C}_{eu}]_{1133}$ & 12.39 & $A^{Z}_e$ & 21.90 & $R_t^{365}$ & 22.44 \\
\hline
$[\mathcal{C}_{uW}]_{33}$ & 22.29 & $A^{Z}_e$ & 1.15 & $\sigma_{Zh}^{365}$ & 22.29 \\
\hline
$[\mathcal{C}_{uB}]_{33}$ & 22.04 & $A^{Z}_e$ & 0.99 & $\sigma_{WW}^{240}$ & 22.04 \\
\hline
$[\mathcal{C}_{le}]_{2211}$ & 2.65 & $A^{Z}_e$ & 21.54 & $R_\mu^{240}$ & 21.54 \\
\hline
$[\mathcal{C}_{le}]_{3311}$ & 2.39 & $A^{Z}_e$ & 21.54 & $R_\tau^{240}$ & 21.54 \\
\hline
$[\mathcal{C}_{le}]_{1122}$ & 2.62 & $A^{Z}_\mu$ & 21.52 & $R_\mu^{240}$ & 21.52 \\
\hline
$[\mathcal{C}_{le}]_{1133}$ & 2.29 & $A^{Z}_e$ & 21.52 & $R_\tau^{240}$ & 21.52 \\
\hline
$[\mathcal{C}_{lu}]_{1133}$ & 11.78 & $A^{Z}_e$ & 20.95 & $R_t^{365}$ & 21.46 \\
\hline
$[\mathcal{C}_{ed}]_{1122}$ & 2.28 & $A^{Z}_e$ & 20.86 & $R_e^{365}$ & 20.86 \\
\hline
$[\mathcal{C}_{qe}]_{1111}$ & 2.30 & $A^{Z}_e$ & 20.82 & $R_e^{365}$ & 20.82 \\
\hline
$[\mathcal{C}_{ed}]_{1111}$ & 2.29 & $A^{Z}_e$ & 20.80 & $R_e^{365}$ & 20.80 \\
\hline
$[\mathcal{C}_{ld}]_{1133}$ & 2.14 & $A^{Z}_e$ & 20.31 & $R_b^{240}$ & 20.31 \\
\hline
$[\mathcal{C}_{ll}]_{1111}$ & 1.06 & $\Gamma_Z$ & 19.42 & $R_e^{365}$ & 19.42 \\
\hline
$[\mathcal{C}_{lu}]_{1111}$ & 3.04 & $A^{Z}_e$ & 19.41 & $R_e^{365}$ & 19.42 \\
\hline
$[\mathcal{C}_{ee}]_{1111}$ & 4.39 & $A^{Z}_e$ & 19.23 & $R_e^{365}$ & 19.25 \\
\hline
$[\mathcal{C}_{qe}]_{2211}$ & 2.38 & $A^{Z}_e$ & 19.20 & $R_e^{365}$ & 19.21 \\
\hline
$[\mathcal{C}_{lq}^{(3)}]_{2233}$ & 17.55 & $A^{Z}_e$ & 7.04 & $\sigma_{WW}^{161}$ & 17.67 \\
\hline
$[\mathcal{C}_{Hq}^{(1)}]_{11}$ & 17.64 & $\Gamma_Z$ & 0.68 & $\sigma_{WW}^{240}$ & 17.64 \\
\hline
$[\mathcal{C}_{le}]_{1111}$ & 3.12 & $A^{Z}_e$ & 15.94 & $R_e^{365}$ & 15.94 \\
\hline
$[\mathcal{C}_{Hq}^{(1)}]_{22}$ & 15.06 & $\Gamma_Z$ & 0.94 & $R_c^{240}$ & 15.06 \\
\hline
$[\mathcal{C}_{lq}^{(1)}]_{1111}$ & 2.04 & $A^{Z}_e$ & 14.58 & $R_e^{365}$ & 14.58 \\
\hline
$[\mathcal{C}_{ld}]_{1122}$ & 2.16 & $A^{Z}_e$ & 13.81 & $R_e^{365}$ & 13.82 \\
\hline
$[\mathcal{C}_{ld}]_{1111}$ & 2.16 & $A^{Z}_e$ & 13.77 & $R_e^{365}$ & 13.77 \\
\hline
$[\mathcal{C}_{Hu}]_{22}$ & 13.17 & $R^{Z}_\mu$ & 1.14 & $R_c^{240}$ & 13.17 \\
\hline
$[\mathcal{C}_{Hu}]_{11}$ & 13.10 & $R^{Z}_\mu$ & 0.95 & $\sigma_{WW}^{240}$ & 13.10 \\
\hline
$[\mathcal{C}_{qe}]_{3322}$ & 13.05 & $A^{Z}_\mu$ & 2.04 & $R_\mu^{163}$ & 13.06 \\
\hline
$[\mathcal{C}_{lq}^{(1)}]_{2233}$ & 12.59 & $A^{Z}_\mu$ & 1.39 & $R_\mu^{163}$ & 12.59 \\
\hline
$[\mathcal{C}_{eu}]_{2233}$ & 12.26 & $A^{Z}_\mu$ & 2.83 & $R_\mu^{163}$ & 12.27 \\
\hline
$[\mathcal{C}_{lu}]_{2233}$ & 11.81 & $A^{Z}_\mu$ & 1.87 & $R_\mu^{163}$ & 11.82 \\
\hline
$\mathcal{C}_{HW}$ & 9.29 & $A^{Z}_e$ & 9.21 & $\sigma_{Zh}^{240}$ & 11.00 \\
\hline
$[\mathcal{C}_{Hd}]_{11}$ & 10.85 & $\Gamma_Z$ & 0.67 & $\sigma_{WW}^{240}$ & 10.85 \\
\hline
$[\mathcal{C}_{Hd}]_{33}$ & 9.64 & $R^{Z}_\mu$ & 0.94 & $R_b^{240}$ & 9.64 \\
\hline
$[\mathcal{C}_{Hq}^{(1)}]_{23}$ & 9.32 & $m_W$ & 0.55 & $\sigma_{WW}^{240}$ & 9.32 \\
\hline
$\mathcal{C}_{HB}$ & 9.16 & $A^{Z}_e$ & 4.71 & $\sigma_{Zh}^{240}$ & 9.31 \\
\hline
$[\mathcal{C}_{qe}]_{3333}$ & 9.08 & $A^{Z}_\tau$ & 2.04 & $R_\tau^{163}$ & 9.09 \\
\hline
$\mathcal{C}_{W}$ & 8.63 & $A^{Z}_e$ & 5.52 & $\sigma_{WW}^{240}$ & 8.97 \\
\hline
$[\mathcal{C}_{lq}^{(1)}]_{3333}$ & 8.82 & $A^{Z}_\tau$ & 1.38 & $R_\tau^{163}$ & 8.82 \\
\hline
$[\mathcal{C}_{qq}^{(1)}]_{3333}$ & 8.69 & $R^{Z}_\mu$ & 2.00 & $R_t^{365}$ & 8.70 \\
\hline
$[\mathcal{C}_{eu}]_{3333}$ & 8.53 & $A^{Z}_\tau$ & 2.83 & $R_\tau^{163}$ & 8.56 \\
\hline
$[\mathcal{C}_{Hd}]_{22}$ & 8.48 & $R^{Z}_\mu$ & 0.67 & $\sigma_{WW}^{240}$ & 8.48 \\
\hline
$[\mathcal{C}_{lu}]_{3333}$ & 8.25 & $A^{Z}_\tau$ & 1.87 & $R_\tau^{163}$ & 8.25 \\
\hline
$[\mathcal{C}_{lq}^{(3)}]_{3333}$ & 7.63 & $A^{Z}_\tau$ & 5.50 & $R_\tau^{163}$ & 8.10 \\
\hline
$[\mathcal{C}_{lq}^{(3)}]_{2211}$ & 6.82 & $A^{Z}_\mu$ & 5.65 & $R_\mu^{163}$ & 7.51 \\
\hline
$[\mathcal{C}_{lq}^{(3)}]_{2222}$ & 6.79 & $A^{Z}_\mu$ & 5.65 & $R_\mu^{163}$ & 7.49 \\
\hline
$\mathcal{C}_{H\square}$ & 6.84 & $m_W$ & 3.96 & $\sigma_{Zh}^{240}$ & 7.02 \\
\hline
$[\mathcal{C}_{qu}^{(1)}]_{3333}$ & 6.72 & $m_W$ & 1.60 & $R_t^{365}$ & 6.73 \\
\hline
$[\mathcal{C}_{qq}^{(3)}]_{1133}$ & 5.91 & $\Gamma_Z$ & 4.51 & $R_c^{240}$ & 6.36 \\
\hline
$[\mathcal{C}_{qq}^{(3)}]_{1111}$ & 5.42 & $\Gamma_Z$ & 5.26 & $R_c^{240}$ & 6.36 \\
\hline
$[\mathcal{C}_{uu}]_{3333}$ & 6.34 & $m_W$ & 2.01 & $R_t^{365}$ & 6.35 \\
\hline
$[\mathcal{C}_{lq}^{(3)}]_{3311}$ & 4.87 & $A^{Z}_\tau$ & 5.61 & $R_\tau^{163}$ & 6.27 \\
\hline
$[\mathcal{C}_{lq}^{(3)}]_{3322}$ & 4.77 & $A^{Z}_\tau$ & 5.60 & $R_\tau^{163}$ & 6.23 \\
\hline
$[\mathcal{C}_{qq}^{(3)}]_{1122}$ & 5.17 & $R^{Z}_\mu$ & 5.05 & $R_\tau^{163}$ & 6.08 \\
\hline
$[\mathcal{C}_{qq}^{(3)}]_{3333}$ & 4.79 & $R^{Z}_\mu$ & 5.17 & $R_b^{240}$ & 5.94 \\
\hline
$[\mathcal{C}_{qq}^{(3)}]_{2222}$ & 4.42 & $R^{Z}_\mu$ & 5.10 & $R_c^{163}$ & 5.70 \\
\hline
$[\mathcal{C}_{qq}^{(3)}]_{2233}$ & 4.72 & $R^{Z}_\mu$ & 4.40 & $R_\tau^{163}$ & 5.43 \\
\hline
$[\mathcal{C}_{uB}]_{23}$ & 5.26 & $A^{Z}_e$ & 0.00 & --- & 5.26 \\
\hline
$[\mathcal{C}_{uW}]_{23}$ & 5.11 & $A^{Z}_e$ & 0.00 & --- & 5.11 \\
\hline
$[\mathcal{C}_{ee}]_{2222}$ & 4.33 & $A^{Z}_\mu$ & 3.96 & $R_\mu^{163}$ & 4.95 \\
\hline
$[\mathcal{C}_{qq}^{(1)}]_{1133}$ & 4.44 & $\Gamma_Z$ & 1.12 & $R_b^{240}$ & 4.45 \\
\hline
$[\mathcal{C}_{ee}]_{3333}$ & 3.01 & $A^{Z}_\tau$ & 3.96 & $R_\tau^{163}$ & 4.26 \\
\hline
$[\mathcal{C}_{Hq}^{(1)}]_{13}$ & 4.20 & $m_W$ & 0.00 & --- & 4.20 \\
\hline
$[\mathcal{C}_{Hud}]_{33}$ & 4.14 & $m_W$ & 0.00 & --- & 4.14 \\
\hline
$[\mathcal{C}_{qu}^{(1)}]_{1133}$ & 4.08 & $\Gamma_Z$ & 1.30 & $R_t^{365}$ & 4.09 \\
\hline
$[\mathcal{C}_{qq}^{(1)}]_{2233}$ & 3.69 & $\Gamma_Z$ & 1.39 & $R_c^{240}$ & 3.71 \\
\hline
$[\mathcal{C}_{eu}]_{2222}$ & 3.21 & $A^{Z}_\mu$ & 2.77 & $R_\mu^{163}$ & 3.58 \\
\hline
$[\mathcal{C}_{eu}]_{2211}$ & 3.21 & $A^{Z}_\mu$ & 2.72 & $R_\mu^{163}$ & 3.56 \\
\hline
$[\mathcal{C}_{uu}]_{2222}$ & 1.17 & $R^{Z}_\mu$ & 3.55 & $R_c^{240}$ & 3.56 \\
\hline
$[\mathcal{C}_{qu}^{(1)}]_{2233}$ & 3.46 & $\Gamma_Z$ & 1.87 & $R_c^{240}$ & 3.53 \\
\hline
$[\mathcal{C}_{qu}^{(1)}]_{3322}$ & 3.47 & $R^{Z}_\mu$ & 1.76 & $R_c^{240}$ & 3.52 \\
\hline
$[\mathcal{C}_{qu}^{(1)}]_{3311}$ & 3.45 & $R^{Z}_\mu$ & 1.58 & $R_b^{240}$ & 3.49 \\
\hline
$[\mathcal{C}_{dB}]_{33}$ & 3.34 & $A^{Z}_e$ & 0.00 & --- & 3.34 \\
\hline
$[\mathcal{C}_{uu}]_{2233}$ & 3.09 & $R^{Z}_\mu$ & 2.23 & $R_c^{240}$ & 3.28 \\
\hline
$[\mathcal{C}_{le}]_{2222}$ & 3.01 & $A^{Z}_\mu$ & 2.39 & $R_\mu^{163}$ & 3.27 \\
\hline
$[\mathcal{C}_{lu}]_{2222}$ & 3.07 & $A^{Z}_\mu$ & 1.90 & $R_\mu^{163}$ & 3.18 \\
\hline
$[\mathcal{C}_{uG}]_{33}$ & 3.17 & $A^{Z}_e$ & 0.00 & --- & 3.17 \\
\hline
$[\mathcal{C}_{uu}]_{1133}$ & 3.08 & $R^{Z}_\mu$ & 1.83 & $R_t^{365}$ & 3.17 \\
\hline
$[\mathcal{C}_{lu}]_{2211}$ & 3.07 & $A^{Z}_\mu$ & 1.76 & $R_\mu^{163}$ & 3.15 \\
\hline
$[\mathcal{C}_{eu}]_{3322}$ & 2.23 & $A^{Z}_\tau$ & 2.77 & $R_\tau^{163}$ & 3.02 \\
\hline
$[\mathcal{C}_{eu}]_{3311}$ & 2.23 & $A^{Z}_\tau$ & 2.72 & $R_\tau^{163}$ & 2.99 \\
\hline
$[\mathcal{C}_{qq}^{(3)}]_{1331}$ & 2.86 & $\Gamma_Z$ & 1.82 & $R_b^{240}$ & 2.97 \\
\hline
$[\mathcal{C}_{uu}]_{1111}$ & 1.16 & $R^{Z}_\mu$ & 2.86 & $R_c^{240}$ & 2.88 \\
\hline
$[\mathcal{C}_{qd}^{(1)}]_{3311}$ & 2.81 & $\Gamma_Z$ & 1.12 & $R_b^{240}$ & 2.83 \\
\hline
$[\mathcal{C}_{qq}^{(1)}]_{2222}$ & 1.47 & $R^{Z}_\mu$ & 2.77 & $R_c^{240}$ & 2.82 \\
\hline
$[\mathcal{C}_{ee}]_{2233}$ & 2.36 & $A^{Z}_\mu$ & 2.35 & $R_\mu^{163}$ & 2.80 \\
\hline
$[\mathcal{C}_{ll}]_{2332}$ & 2.28 & $A^{Z}_\mu$ & 2.37 & $R_\mu^{163}$ & 2.76 \\
\hline
$[\mathcal{C}_{qq}^{(1)}]_{1111}$ & 2.00 & $\Gamma_Z$ & 2.53 & $R_c^{240}$ & 2.75 \\
\hline
$[\mathcal{C}_{le}]_{3333}$ & 2.07 & $A^{Z}_\tau$ & 2.39 & $R_\tau^{163}$ & 2.67 \\
\hline
$[\mathcal{C}_{le}]_{3322}$ & 2.38 & $A^{Z}_\mu$ & 2.08 & $R_\mu^{163}$ & 2.67 \\
\hline
$[\mathcal{C}_{le}]_{2233}$ & 2.32 & $A^{Z}_\mu$ & 2.08 & $R_\tau^{163}$ & 2.63 \\
\hline
$[\mathcal{C}_{qe}]_{2222}$ & 2.34 & $A^{Z}_\mu$ & 2.00 & $R_\mu^{163}$ & 2.60 \\
\hline
$[\mathcal{C}_{ud}^{(1)}]_{3311}$ & 2.54 & $\Gamma_Z$ & 1.29 & $R_t^{365}$ & 2.58 \\
\hline
$[\mathcal{C}_{qd}^{(1)}]_{3333}$ & 2.54 & $R^{Z}_\mu$ & 1.12 & $R_b^{240}$ & 2.56 \\
\hline
$[\mathcal{C}_{ed}]_{2233}$ & 2.24 & $A^{Z}_\mu$ & 2.00 & $R_\mu^{163}$ & 2.53 \\
\hline
$[\mathcal{C}_{qe}]_{1122}$ & 2.27 & $A^{Z}_\mu$ & 1.92 & $R_\mu^{163}$ & 2.51 \\
\hline
$[\mathcal{C}_{ed}]_{2211}$ & 2.25 & $A^{Z}_\mu$ & 1.92 & $R_\mu^{163}$ & 2.50 \\
\hline
$[\mathcal{C}_{ed}]_{2222}$ & 2.25 & $A^{Z}_\mu$ & 1.92 & $R_\mu^{163}$ & 2.50 \\
\hline
$[\mathcal{C}_{ud}^{(1)}]_{3333}$ & 2.26 & $R^{Z}_\mu$ & 1.86 & $R_b^{240}$ & 2.48 \\
\hline
$[\mathcal{C}_{ll}]_{2233}$ & 2.31 & $A^{Z}_\mu$ & 1.58 & $R_\tau^{163}$ & 2.43 \\
\hline
$[\mathcal{C}_{qu}^{(1)}]_{2222}$ & 0.88 & $\Gamma_Z$ & 2.42 & $R_c^{240}$ & 2.43 \\
\hline
$[\mathcal{C}_{lu}]_{3322}$ & 2.13 & $A^{Z}_\tau$ & 1.90 & $R_\tau^{163}$ & 2.41 \\
\hline
$[\mathcal{C}_{uu}]_{1122}$ & 1.06 & $R^{Z}_\mu$ & 2.35 & $R_c^{163}$ & 2.37 \\
\hline
$[\mathcal{C}_{qq}^{(3)}]_{2332}$ & 2.16 & $\Gamma_Z$ & 1.75 & $R_b^{240}$ & 2.36 \\
\hline
$[\mathcal{C}_{uB}]_{13}$ & 2.36 & $A^{Z}_e$ & 0.00 & --- & 2.36 \\
\hline
$[\mathcal{C}_{lu}]_{3311}$ & 2.13 & $A^{Z}_\tau$ & 1.76 & $R_\tau^{163}$ & 2.35 \\
\hline
$[\mathcal{C}_{ld}]_{2233}$ & 2.17 & $A^{Z}_\mu$ & 1.48 & $R_b^{240}$ & 2.28 \\
\hline
$[\mathcal{C}_{qd}^{(1)}]_{3322}$ & 2.22 & $R^{Z}_\mu$ & 1.12 & $R_b^{240}$ & 2.26 \\
\hline
$[\mathcal{C}_{uW}]_{13}$ & 2.26 & $A^{Z}_e$ & 0.00 & --- & 2.26 \\
\hline
$[\mathcal{C}_{lq}^{(1)}]_{2222}$ & 2.14 & $A^{Z}_\mu$ & 1.44 & $R_c^{240}$ & 2.25 \\
\hline
$[\mathcal{C}_{ld}]_{2211}$ & 2.18 & $A^{Z}_\mu$ & 1.25 & $R_\mu^{163}$ & 2.24 \\
\hline
$[\mathcal{C}_{ld}]_{2222}$ & 2.18 & $A^{Z}_\mu$ & 1.25 & $R_\mu^{163}$ & 2.24 \\
\hline
$[\mathcal{C}_{qq}^{(3)}]_{1221}$ & 1.81 & $R^{Z}_\mu$ & 1.92 & $R_\tau^{163}$ & 2.22 \\
\hline
$[\mathcal{C}_{ll}]_{2222}$ & 1.05 & $\Gamma_Z$ & 2.16 & $R_\mu^{163}$ & 2.19 \\
\hline
$[\mathcal{C}_{qe}]_{2233}$ & 1.63 & $A^{Z}_\tau$ & 2.00 & $R_\tau^{163}$ & 2.19 \\
\hline
$[\mathcal{C}_{qq}^{(1)}]_{1221}$ & 1.62 & $R^{Z}_\mu$ & 1.98 & $R_\tau^{163}$ & 2.17 \\
\hline
$[\mathcal{C}_{ed}]_{3333}$ & 1.56 & $A^{Z}_\tau$ & 2.00 & $R_\tau^{163}$ & 2.16 \\
\hline
$[\mathcal{C}_{Hq}^{(3)}]_{23}$ & 2.14 & $m_W$ & 0.38 & $\sigma_{Zh}^{240}$ & 2.14 \\
\hline
$[\mathcal{C}_{qu}^{(1)}]_{2333}$ & 2.12 & $m_W$ & 0.00 & --- & 2.12 \\
\hline
$[\mathcal{C}_{lq}^{(1)}]_{2211}$ & 2.07 & $A^{Z}_\mu$ & 1.19 & $R_\mu^{163}$ & 2.12 \\
\hline
$[\mathcal{C}_{qe}]_{1133}$ & 1.60 & $A^{Z}_\tau$ & 1.92 & $R_\tau^{163}$ & 2.12 \\
\hline
$[\mathcal{C}_{dd}]_{3333}$ & 0.62 & $R^{Z}_\mu$ & 2.11 & $R_b^{240}$ & 2.11 \\
\hline
$[\mathcal{C}_{ll}]_{3333}$ & 1.04 & $\Gamma_Z$ & 2.07 & $R_\tau^{163}$ & 2.11 \\
\hline
$[\mathcal{C}_{ed}]_{3311}$ & 1.57 & $A^{Z}_\tau$ & 1.92 & $R_\tau^{163}$ & 2.10 \\
\hline
$[\mathcal{C}_{ed}]_{3322}$ & 1.57 & $A^{Z}_\tau$ & 1.92 & $R_\tau^{163}$ & 2.10 \\
\hline
$[\mathcal{C}_{ud}^{(1)}]_{3322}$ & 1.99 & $R^{Z}_\mu$ & 1.29 & $R_t^{365}$ & 2.07 \\
\hline
$[\mathcal{C}_{eW}]_{33}$ & 2.07 & $A^{Z}_e$ & 0.00 & --- & 2.07 \\
\hline
$[\mathcal{C}_{ud}^{(1)}]_{2233}$ & 0.75 & $R^{Z}_\mu$ & 2.01 & $R_b^{240}$ & 2.02 \\
\hline
$[\mathcal{C}_{ud}^{(1)}]_{1133}$ & 0.75 & $R^{Z}_\mu$ & 1.89 & $R_b^{240}$ & 1.90 \\
\hline
$[\mathcal{C}_{qq}^{(1)}]_{2332}$ & 1.44 & $\Gamma_Z$ & 1.71 & $R_\tau^{163}$ & 1.90 \\
\hline
$[\mathcal{C}_{qu}^{(1)}]_{2211}$ & 0.89 & $\Gamma_Z$ & 1.84 & $R_c^{240}$ & 1.86 \\
\hline
$[\mathcal{C}_{qu}^{(1)}]_{1111}$ & 1.15 & $\Gamma_Z$ & 1.78 & $R_c^{240}$ & 1.85 \\
\hline
$[\mathcal{C}_{qq}^{(1)}]_{1331}$ & 1.26 & $R^{Z}_\mu$ & 1.74 & $R_c^{240}$ & 1.85 \\
\hline
$[\mathcal{C}_{ud}^{(1)}]_{1111}$ & 0.82 & $\Gamma_Z$ & 1.77 & $R_c^{240}$ & 1.79 \\
\hline
$[\mathcal{C}_{ud}^{(1)}]_{1122}$ & 0.70 & $R^{Z}_\mu$ & 1.77 & $R_c^{240}$ & 1.79 \\
\hline
$[\mathcal{C}_{ld}]_{3333}$ & 1.51 & $A^{Z}_\tau$ & 1.48 & $R_b^{240}$ & 1.77 \\
\hline
$[\mathcal{C}_{qu}^{(1)}]_{1122}$ & 1.15 & $\Gamma_Z$ & 1.66 & $R_c^{163}$ & 1.75 \\
\hline
$[\mathcal{C}_{lq}^{(1)}]_{3322}$ & 1.49 & $A^{Z}_\tau$ & 1.44 & $R_c^{240}$ & 1.75 \\
\hline
$[\mathcal{C}_{ud}^{(1)}]_{2211}$ & 0.82 & $\Gamma_Z$ & 1.66 & $R_c^{163}$ & 1.69 \\
\hline
$[\mathcal{C}_{ud}^{(1)}]_{2222}$ & 0.70 & $R^{Z}_\mu$ & 1.66 & $R_c^{163}$ & 1.68 \\
\hline
$[\mathcal{C}_{ld}]_{3311}$ & 1.52 & $A^{Z}_\tau$ & 1.25 & $R_\tau^{163}$ & 1.67 \\
\hline
$[\mathcal{C}_{ld}]_{3322}$ & 1.51 & $A^{Z}_\tau$ & 1.25 & $R_\tau^{163}$ & 1.67 \\
\hline
$[\mathcal{C}_{uu}]_{2332}$ & 1.47 & $R^{Z}_\mu$ & 1.22 & $R_c^{240}$ & 1.62 \\
\hline
$[\mathcal{C}_{eB}]_{33}$ & 1.61 & $A^{Z}_e$ & 0.00 & --- & 1.61 \\
\hline
$[\mathcal{C}_{lq}^{(1)}]_{3311}$ & 1.47 & $A^{Z}_\tau$ & 1.19 & $R_\tau^{163}$ & 1.60 \\
\hline
$[\mathcal{C}_{uB}]_{22}$ & 1.57 & $A^{Z}_e$ & 0.00 & --- & 1.57 \\
\hline
$[\mathcal{C}_{qd}^{(1)}]_{2233}$ & 0.62 & $\Gamma_Z$ & 1.53 & $R_b^{240}$ & 1.54 \\
\hline
$[\mathcal{C}_{uu}]_{1331}$ & 1.46 & $R^{Z}_\mu$ & 1.00 & $R_t^{365}$ & 1.54 \\
\hline
$[\mathcal{C}_{dd}]_{1111}$ & 0.70 & $\Gamma_Z$ & 1.43 & $R_c^{240}$ & 1.45 \\
\hline
$[\mathcal{C}_{dd}]_{2222}$ & 0.54 & $R^{Z}_\mu$ & 1.43 & $R_c^{240}$ & 1.44 \\
\hline
$[\mathcal{C}_{uW}]_{22}$ & 1.41 & $A^{Z}_e$ & 0.00 & --- & 1.41 \\
\hline
$[\mathcal{C}_{dW}]_{33}$ & 1.38 & $A^{Z}_e$ & 0.00 & --- & 1.38 \\
\hline
$[\mathcal{C}_{qd}^{(1)}]_{1133}$ & 0.81 & $\Gamma_Z$ & 1.34 & $R_b^{240}$ & 1.38 \\
\hline
$[\mathcal{C}_{dd}]_{1133}$ & 0.58 & $\Gamma_Z$ & 1.33 & $R_b^{240}$ & 1.35 \\
\hline
$[\mathcal{C}_{dd}]_{2233}$ & 0.50 & $R^{Z}_\mu$ & 1.34 & $R_b^{240}$ & 1.34 \\
\hline
$[\mathcal{C}_{qd}^{(1)}]_{2222}$ & 0.68 & $\Gamma_Z$ & 1.30 & $R_c^{240}$ & 1.32 \\
\hline
$[\mathcal{C}_{qd}^{(1)}]_{1111}$ & 0.86 & $\Gamma_Z$ & 1.26 & $R_c^{240}$ & 1.32 \\
\hline
$[\mathcal{C}_{qd}^{(1)}]_{2211}$ & 0.58 & $R^{Z}_\mu$ & 1.30 & $R_c^{240}$ & 1.31 \\
\hline
$[\mathcal{C}_{qd}^{(1)}]_{1122}$ & 0.77 & $\Gamma_Z$ & 1.26 & $R_c^{240}$ & 1.30 \\
\hline
$[\mathcal{C}_{uu}]_{1221}$ & 0.47 & $R^{Z}_\mu$ & 1.28 & $R_\mu^{240}$ & 1.28 \\
\hline
$[\mathcal{C}_{qq}^{(1)}]_{1122}$ & 0.37 & $R^{Z}_c$ & 1.26 & $R_c^{240}$ & 1.27 \\
\hline
$[\mathcal{C}_{dd}]_{1122}$ & 0.53 & $R^{Z}_\mu$ & 1.25 & $R_c^{240}$ & 1.26 \\
\hline
$[\mathcal{C}_{Hq}^{(3)}]_{13}$ & 1.10 & $m_W$ & 0.35 & $\sigma_{Zh}^{240}$ & 1.10 \\
\hline
$[\mathcal{C}_{qu}^{(1)}]_{1333}$ & 1.08 & $m_W$ & 0.00 & --- & 1.08 \\
\hline
$\mathcal{C}_{H}$ & 0.39 & $m_W$ & 0.99 & $\sigma_{Zh}^{240}$ & 0.99 \\
\hline
\end{longtable}
\normalsize

\subsection{SILH Basis}
\label{app:SILH_bounds}
We report here the projected sensitivity to universal operators in the SILH basis as well as a select number of non-universal operators which are defined differently to the Warsaw basis. Concretely, in addition to the operators introduced in Eqs. \ref{eq:OT}-\ref{eq:2W} and \ref{eq:opHW} -\ref{eq:op3W}, we focus on 

\begin{equation}
\begin{array}{ccccc}
\mathcal{O}_{\gamma W} = Q_{HW}, &\hspace{0.5cm} \mathcal{O}_{\gamma B} = Q_{HB}, &\hspace{0.5cm} \mathcal{O}_{3G} = Q_{3G}, &\hspace{0.5cm} \mathcal{O}_{H} = -\frac{1}{2}Q_{H\square}, &\hspace{0.5cm} \mathcal{O}_{6} = Q_{H} ,
\end{array}
\end{equation}
which are just slightly modified and renamed with respect to their Warsaw Basis counterparts $Q_i$, as well as 
\begin{equation}
    \mathcal{O}_{2G} = -\frac{1}{2}(D^{\mu}G^{A}_{\mu\nu})(D_{\rho}G^{A,\rho\nu}). 
\end{equation}
In addition to those universal contributions, we also define the following operators modifying the electroweak vertices to third family quarks: 
\begin{align}
    \mathcal{O}_{qD}^{(1)} &= (\bar{q}_L^3\gamma^{\mu}q_L^3)\partial^\nu B_{\mu\nu}\\
    \mathcal{O}_{qD}^{(3)} &= (\bar{q}_L^3\gamma^{\mu}\tau^{I}q_L^3)D^\nu W_{\mu\nu}^{I}\\
    \mathcal{O}_{tD} &= (\bar{t}_R\gamma^{\mu}t_R)\partial^\nu B_{\mu\nu}\,.
\end{align}

\small
\begin{longtable}{|c|c|c|c|c|c|}
\caption{Single Operator $2\sigma$ confidence intervals in the  basis exceding \SI{1}{\tera\electronvolt} from On- and Off-Pole observables, as well as their combined projected sensitivity. The Wilson Coefficients are renormalized at $\Lambda=\SI{1}{\tera\electronvolt}$. We also report the most constraining measurement for each coefficient for On- and Off-Pole Observables.} \\
\hline
\toprule
$\mathcal{S}_{\rm{i}}$ & $\Lambda^{\rm{eff}}_{\rm{i}}$ on-pole & Obs. on-pole & $\Lambda^{\rm{eff}}_{\rm{i}}$ off-pole & Obs. off-pole & $\Lambda^{\rm{eff}}_{\rm{i}}$ Combined \\
\hline
\midrule
\endfirsthead
\caption[]{$2\sigma$ CL in the SILH basis. Coefficients are renormalized at $\Lambda=\SI{1}{\tera\electronvolt}$} \\
\hline
\toprule
$\mathcal{S}_{\rm{i}}$ & $\Lambda^{\rm{eff}}_{\rm{i}}$ on-pole & Obs. on-pole & $\Lambda^{\rm{eff}}_{\rm{i}}$ off-pole & Obs. off-pole & $\Lambda^{\rm{eff}}_{\rm{i}}$ Combined \\
\hline
\midrule
\endhead
\midrule
\multicolumn{6}{r}{Continued on next page} \\
\hline
\midrule
\endfoot
\bottomrule
\endlastfoot
$\mathcal{S}_{T}$ & 80.62 & $m_W$ & 6.41 & $\sigma_{WW}^{240}$ & 80.62 \\
\hline
$\mathcal{S}_{2W}$ & 21.88 & $A^{Z}_e$ & 25.01 & $R_\tau^{240}$ & 28.06 \\
\hline
$\mathcal{S}_{qD}^{(3)}$ & 13.02 & $R^{Z}_\mu$ & 26.52 & $R_b^{240}$ & 26.90 \\
\hline
$\mathcal{S}_{B + W}$ & 26.57 & $A^{Z}_e$ & 4.25 & $\sigma_{Zh}^{365}$ & 26.58 \\
\hline
$\mathcal{S}_{2B}$ & 23.25 & $A^{Z}_e$ & 13.13 & $R_e^{365}$ & 23.82 \\
\hline
$\mathcal{S}_{qD}^{(1)}$ & 16.18 & $A^{Z}_e$ & 16.82 & $R_t^{365}$ & 19.63 \\
\hline
$\mathcal{S}_{tD}$ & 14.49 & $A^{Z}_e$ & 15.73 & $R_t^{365}$ & 18.01 \\
\hline
$\mathcal{S}_{\gamma W}$ & 9.29 & $A^{Z}_e$ & 9.21 & $\sigma_{Zh}^{240}$ & 11.00 \\
\hline
$\mathcal{S}_{\gamma B}$ & 9.16 & $A^{Z}_e$ & 4.71 & $\sigma_{Zh}^{240}$ & 9.31 \\
\hline
$\mathcal{S}_{3W}$ & 8.63 & $A^{Z}_e$ & 5.52 & $\sigma_{WW}^{240}$ & 8.97 \\
\hline
$\mathcal{S}_{HW}$ & 5.27 & $m_W$ & 5.22 & $\sigma_{Zh}^{365}$ & 6.24 \\
\hline
$\mathcal{S}_{HB}$ & 5.28 & $m_W$ & 4.11 & $\sigma_{WW}^{240}$ & 5.71 \\
\hline
$\mathcal{S}_{H}$ & 4.84 & $m_W$ & 2.80 & $\sigma_{Zh}^{240}$ & 4.97 \\
\hline
$\mathcal{S}_{2G}$ & 3.87 & $m_W$ & 3.49 & $R_\mu^{163}$ & 4.40 \\
\hline
$\mathcal{S}_{B - W}$ & 3.38 & $A^{Z}_e$ & 3.23 & $\sigma_{Zh}^{365}$ & 3.94 \\
\hline
$\mathcal{S}_{6}$ & 0.39 & $m_W$ & 0.99 & $\sigma_{Zh}^{240}$ & 0.99 \\
\hline
$\mathcal{S}_{3G}$ & 0.53 & $m_W$ & 0.00 & --- & 0.53 \\
\hline
$\mathcal{S}_{g}$ & 0.47 & $m_W$ & 0.00 & --- & 0.47 \\
\hline
\end{longtable}
\normalsize

\bibliographystyle{JHEP}
\bibliography{main}

\providecommand{\href}[2]{#2}\begingroup\raggedright\begin{thebibliography}{10}

\bibitem{CERN-ESU-015}
{\it {2020 Update of the European Strategy for Particle Physics (Brochure)}},  tech. rep., Geneva, 2020.

\bibitem{osti_2368847}
H.~Murayama, S.~Asai, K.~Heeger, A.~Ballarino, T.~Bose, K.~Cranmer, F.-Y. Cyr-Racine, S.~Demers, C.~Geddes, Y.~Gershtein, B.~Heinemann, J.~Hewett, P.~Huber, K.~Mahn, R.~Mandelbaum, J.~Maricic, P.~Merkel, C.~Monahan, P.~Onyisi, M.~Palmer, T.~Raubenheimer, M.~Sanchez, R.~Schnee, S.~Seidel, S.-H. Seo, J.~Thaler, C.~Touramanis, A.~Vieregg, A.~Weinstein, L.~Winslow, T.-T. Yu, and R.~Zwaska, {\it Exploring the quantum universe: Pathways to innovation and discovery in particle physics}, .

\bibitem{Blondel:2024mry}
A.~Blondel, C.~Grojean, P.~Janot, and G.~Wilkinson, {\it {Higgs Factory options for CERN: A comparative study}},  \href{http://arxiv.org/abs/2412.13130}{{\tt arXiv:2412.13130}}.

\bibitem{Allwicher:2023aql}
L.~Allwicher, G.~Isidori, J.~M. Lizana, N.~Selimovic, and B.~A. Stefanek, {\it {Third-family quark-lepton Unification and electroweak precision tests}},  {\em JHEP} {\bf 05} (2023) 179, [\href{http://arxiv.org/abs/2302.11584}{{\tt arXiv:2302.11584}}].

\bibitem{Allwicher:2023shc}
L.~Allwicher, C.~Cornella, G.~Isidori, and B.~A. Stefanek, {\it {New physics in the third generation. A comprehensive SMEFT analysis and future prospects}},  {\em JHEP} {\bf 03} (2024) 049, [\href{http://arxiv.org/abs/2311.00020}{{\tt arXiv:2311.00020}}].

\bibitem{Stefanek:2024kds}
B.~A. Stefanek, {\it {Non-universal probes of composite Higgs models: new bounds and prospects for FCC-ee}},  {\em JHEP} {\bf 09} (2024) 103, [\href{http://arxiv.org/abs/2407.09593}{{\tt arXiv:2407.09593}}].

\bibitem{Allwicher:2024sso}
L.~Allwicher, M.~McCullough, and S.~Renner, {\it {New Physics at Tera-$Z$: Precision Renormalised}},  \href{http://arxiv.org/abs/2408.03992}{{\tt arXiv:2408.03992}}.

\bibitem{Erdelyi:2024sls}
B.~A. Erdelyi, R.~Gr\"ober, and N.~Selimovic, {\it {How large can the Light Quark Yukawa couplings be?}},  \href{http://arxiv.org/abs/2410.08272}{{\tt arXiv:2410.08272}}.

\bibitem{Gargalionis:2024jaw}
J.~Gargalionis, J.~Quevillon, P.~N.~H. Vuong, and T.~You, {\it {Linear Standard Model extensions in the SMEFT at one loop and Tera-Z}},  \href{http://arxiv.org/abs/2412.01759}{{\tt arXiv:2412.01759}}.

\bibitem{Davighi:2024syj}
J.~Davighi, {\it {In Search of an Invisible $Z^\prime$}},  \href{http://arxiv.org/abs/2412.07694}{{\tt arXiv:2412.07694}}.

\bibitem{FCC:2018evy}
{\bf FCC} Collaboration, A.~Abada et~al., {\it {FCC-ee: The Lepton Collider}: {Future Circular Collider Conceptual Design Report Volume 2}},  {\em Eur. Phys. J. ST} {\bf 228} (2019), no.~2 261--623.

\bibitem{Bernardi:2022hny}
G.~Bernardi et~al., {\it {The Future Circular Collider: a Summary for the US 2021 Snowmass Process}},  \href{http://arxiv.org/abs/2203.06520}{{\tt arXiv:2203.06520}}.

\bibitem{CEPCStudyGroup:2023quu}
{\bf CEPC Study Group} Collaboration, W.~Abdallah et~al., {\it {CEPC Technical Design Report: Accelerator}},  {\em Radiat. Detect. Technol. Methods} {\bf 8} (2024), no.~1 1--1105, [\href{http://arxiv.org/abs/2312.14363}{{\tt arXiv:2312.14363}}].

\bibitem{Peskin:1991sw}
M.~E. Peskin and T.~Takeuchi, {\it {Estimation of oblique electroweak corrections}},  {\em Phys. Rev. D} {\bf 46} (1992) 381--409.

\bibitem{Farina:2016rws}
M.~Farina, G.~Panico, D.~Pappadopulo, J.~T. Ruderman, R.~Torre, and A.~Wulzer, {\it {Energy helps accuracy: electroweak precision tests at hadron colliders}},  {\em Phys. Lett. B} {\bf 772} (2017) 210--215, [\href{http://arxiv.org/abs/1609.08157}{{\tt arXiv:1609.08157}}].

\bibitem{Dawson:2022bxd}
S.~Dawson and P.~P. Giardino, {\it {Flavorful electroweak precision observables in the Standard Model effective field theory}},  {\em Phys. Rev. D} {\bf 105} (2022), no.~7 073006, [\href{http://arxiv.org/abs/2201.09887}{{\tt arXiv:2201.09887}}].

\bibitem{Bellafronte:2023amz}
L.~Bellafronte, S.~Dawson, and P.~P. Giardino, {\it {The importance of flavor in SMEFT Electroweak Precision Fits}},  {\em JHEP} {\bf 05} (2023) 208, [\href{http://arxiv.org/abs/2304.00029}{{\tt arXiv:2304.00029}}].

\bibitem{Freitas:2019bre}
A.~Freitas et~al., {\it {Theoretical uncertainties for electroweak and Higgs-boson precision measurements at FCC-ee}},  \href{http://arxiv.org/abs/1906.05379}{{\tt arXiv:1906.05379}}.

\bibitem{Maksymyk:1993zm}
I.~Maksymyk, C.~P. Burgess, and D.~London, {\it {Beyond S, T and U}},  {\em Phys. Rev. D} {\bf 50} (1994) 529--535, [\href{http://arxiv.org/abs/hep-ph/9306267}{{\tt hep-ph/9306267}}].

\bibitem{Barbieri:2004qk}
R.~Barbieri, A.~Pomarol, R.~Rattazzi, and A.~Strumia, {\it {Electroweak symmetry breaking after LEP-1 and LEP-2}},  {\em Nucl. Phys. B} {\bf 703} (2004) 127--146, [\href{http://arxiv.org/abs/hep-ph/0405040}{{\tt hep-ph/0405040}}].

\bibitem{Greljo:2024ytg}
A.~Greljo, H.~Tiblom, and A.~Valenti, {\it {New Physics Through Flavor Tagging at FCC-ee}},  \href{http://arxiv.org/abs/2411.02485}{{\tt arXiv:2411.02485}}.

\bibitem{Durieux:2022hbu}
G.~Durieux, M.~McCullough, and E.~Salvioni, {\it {Charting the Higgs self-coupling boundaries}},  {\em JHEP} {\bf 12} (2022) 148, [\href{http://arxiv.org/abs/2209.00666}{{\tt arXiv:2209.00666}}]. [Erratum: JHEP 02, 165 (2023)].

\bibitem{Banta:2021dek}
I.~Banta, T.~Cohen, N.~Craig, X.~Lu, and D.~Sutherland, {\it {Non-decoupling new particles}},  {\em JHEP} {\bf 02} (2022) 029, [\href{http://arxiv.org/abs/2110.02967}{{\tt arXiv:2110.02967}}].

\bibitem{Crawford:2024nun}
G.~Crawford and D.~Sutherland, {\it {Non-decoupling scalars at future colliders}},  \href{http://arxiv.org/abs/2409.18177}{{\tt arXiv:2409.18177}}.

\bibitem{DeBlas:2019qco}
J.~De~Blas, G.~Durieux, C.~Grojean, J.~Gu, and A.~Paul, {\it {On the future of Higgs, electroweak and diboson measurements at lepton colliders}},  {\em JHEP} {\bf 12} (2019) 117, [\href{http://arxiv.org/abs/1907.04311}{{\tt arXiv:1907.04311}}].

\bibitem{Blondel:2018mad}
A.~Blondel et~al., {\it {Standard model theory for the FCC-ee Tera-Z stage}},  in {\em {Mini Workshop on Precision EW and QCD Calculations for the FCC Studies : Methods and Techniques}}, vol.~3/2019 of {\em CERN Yellow Reports: Monographs}, (Geneva), CERN, 9, 2018.
\newblock \href{http://arxiv.org/abs/1809.01830}{{\tt arXiv:1809.01830}}.

\bibitem{deBlas:2022ofj}
J.~de~Blas, Y.~Du, C.~Grojean, J.~Gu, V.~Miralles, M.~E. Peskin, J.~Tian, M.~Vos, and E.~Vryonidou, {\it {Global SMEFT Fits at Future Colliders}},  in {\em {Snowmass 2021}}, 6, 2022.
\newblock \href{http://arxiv.org/abs/2206.08326}{{\tt arXiv:2206.08326}}.

\bibitem{Asteriadis:2024xts}
K.~Asteriadis, S.~Dawson, P.~P. Giardino, and R.~Szafron, {\it {The $e^+ e^- \rightarrow Z H$ Process in the SMEFT Beyond Leading Order}},  \href{http://arxiv.org/abs/2409.11466}{{\tt arXiv:2409.11466}}.

\bibitem{Biekotter:2025nln}
A.~Biek\"otter and B.~D. Pecjak, {\it {Analytic results for electroweak precision observables at NLO in SMEFT}},  \href{http://arxiv.org/abs/2503.07724}{{\tt arXiv:2503.07724}}.

\bibitem{Biekotter:2023xle}
A.~Biek\"otter, B.~D. Pecjak, D.~J. Scott, and T.~Smith, {\it {Electroweak input schemes and universal corrections in SMEFT}},  {\em JHEP} {\bf 07} (2023) 115, [\href{http://arxiv.org/abs/2305.03763}{{\tt arXiv:2305.03763}}].

\bibitem{Davighi:2023evx}
J.~Davighi and B.~A. Stefanek, {\it {Deconstructed hypercharge: a natural model of flavour}},  {\em JHEP} {\bf 11} (2023) 100, [\href{http://arxiv.org/abs/2305.16280}{{\tt arXiv:2305.16280}}].

\bibitem{Ellis:2020unq}
J.~Ellis, M.~Madigan, K.~Mimasu, V.~Sanz, and T.~You, {\it {Top, Higgs, Diboson and Electroweak Fit to the Standard Model Effective Field Theory}},  {\em JHEP} {\bf 04} (2021) 279, [\href{http://arxiv.org/abs/2012.02779}{{\tt arXiv:2012.02779}}].

\bibitem{Benedikt:2928193}
M.~Benedikt, W.~Bartmann, J.-P. Burnet, C.~Carli, A.~Chance, P.~Craievich, M.~Giovannozzi, C.~Grojean, J.~Gutleber, K.~Hanke, A.~Henriques, P.~Janot, C.~Lourenco, M.~Mangano, T.~Otto, J.~H. Poole, S.~Rajagopalan, T.~Raubenheimer, E.~Todesco, L.~Ulrici, T.~P. Watson, G.~Wilkinson, F.~Zimmermann, and B.~Auchmann, {\it {Future Circular Collider Feasibility Study Report Volume 1: Physics and Experiments}},  tech. rep., CERN, Geneva, 2025.

\bibitem{Giudice:2007fh}
G.~F. Giudice, C.~Grojean, A.~Pomarol, and R.~Rattazzi, {\it {The Strongly-Interacting Light Higgs}},  {\em JHEP} {\bf 06} (2007) 045, [\href{http://arxiv.org/abs/hep-ph/0703164}{{\tt hep-ph/0703164}}].

\bibitem{Contino:2013kra}
R.~Contino, M.~Ghezzi, C.~Grojean, M.~Muhlleitner, and M.~Spira, {\it {Effective Lagrangian for a light Higgs-like scalar}},  {\em JHEP} {\bf 07} (2013) 035, [\href{http://arxiv.org/abs/1303.3876}{{\tt arXiv:1303.3876}}].

\bibitem{Bissmann:2020mfi}
S.~Bi\ss{}mann, C.~Grunwald, G.~Hiller, and K.~Kr\"oninger, {\it {Top and Beauty synergies in SMEFT-fits at present and future colliders}},  {\em JHEP} {\bf 06} (2021) 010, [\href{http://arxiv.org/abs/2012.10456}{{\tt arXiv:2012.10456}}].

\bibitem{Ethier:2021bye}
{\bf SMEFiT} Collaboration, J.~J. Ethier, G.~Magni, F.~Maltoni, L.~Mantani, E.~R. Nocera, J.~Rojo, E.~Slade, E.~Vryonidou, and C.~Zhang, {\it {Combined SMEFT interpretation of Higgs, diboson, and top quark data from the LHC}},  {\em JHEP} {\bf 11} (2021) 089, [\href{http://arxiv.org/abs/2105.00006}{{\tt arXiv:2105.00006}}].

\bibitem{Grunwald:2023nli}
C.~Grunwald, G.~Hiller, K.~Kr\"oninger, and L.~Nollen, {\it {More synergies from beauty, top, Z and Drell-Yan measurements in SMEFT}},  {\em JHEP} {\bf 11} (2023) 110, [\href{http://arxiv.org/abs/2304.12837}{{\tt arXiv:2304.12837}}].

\bibitem{Garosi:2023yxg}
F.~Garosi, D.~Marzocca, A.~R. S\'anchez, and A.~Stanzione, {\it {Indirect constraints on top quark operators from a global SMEFT analysis}},  {\em JHEP} {\bf 12} (2023) 129, [\href{http://arxiv.org/abs/2310.00047}{{\tt arXiv:2310.00047}}].

\bibitem{Bartocci:2023nvp}
R.~Bartocci, A.~Biek\"otter, and T.~Hurth, {\it {A global analysis of the SMEFT under the minimal MFV assumption}},  {\em JHEP} {\bf 05} (2024) 074, [\href{http://arxiv.org/abs/2311.04963}{{\tt arXiv:2311.04963}}].

\bibitem{Celada:2024mcf}
E.~Celada, T.~Giani, J.~ter Hoeve, L.~Mantani, J.~Rojo, A.~N. Rossia, M.~O.~A. Thomas, and E.~Vryonidou, {\it {Mapping the SMEFT at high-energy colliders: from LEP and the (HL-)LHC to the FCC-ee}},  {\em JHEP} {\bf 09} (2024) 091, [\href{http://arxiv.org/abs/2404.12809}{{\tt arXiv:2404.12809}}].

\bibitem{Bartocci:2024fmm}
R.~Bartocci, A.~Biek\"otter, and T.~Hurth, {\it {Renormalisation group evolution effects on global SMEFT analyses}},  \href{http://arxiv.org/abs/2412.09674}{{\tt arXiv:2412.09674}}.

\bibitem{Grzadkowski:2010es}
B.~Grzadkowski, M.~Iskrzynski, M.~Misiak, and J.~Rosiek, {\it {Dimension-Six Terms in the Standard Model Lagrangian}},  {\em JHEP} {\bf 10} (2010) 085, [\href{http://arxiv.org/abs/1008.4884}{{\tt arXiv:1008.4884}}].

\bibitem{McCullough:2013rea}
M.~McCullough, {\it {An Indirect Model-Dependent Probe of the Higgs Self-Coupling}},  {\em Phys. Rev. D} {\bf 90} (2014), no.~1 015001, [\href{http://arxiv.org/abs/1312.3322}{{\tt arXiv:1312.3322}}]. [Erratum: Phys.Rev.D 92, 039903 (2015)].

\bibitem{Alonso:2013hga}
R.~Alonso, E.~E. Jenkins, A.~V. Manohar, and M.~Trott, {\it {Renormalization Group Evolution of the Standard Model Dimension Six Operators III: Gauge Coupling Dependence and Phenomenology}},  {\em JHEP} {\bf 04} (2014) 159, [\href{http://arxiv.org/abs/1312.2014}{{\tt arXiv:1312.2014}}].

\bibitem{vanderBij:1985ww}
J.~J. van~der Bij, {\it {Does Low-energy Physics Depend on the Potential of a Heavy Higgs Particle?}},  {\em Nucl. Phys. B} {\bf 267} (1986) 557--565.

\bibitem{Kribs:2017znd}
G.~D. Kribs, A.~Maier, H.~Rzehak, M.~Spannowsky, and P.~Waite, {\it {Electroweak oblique parameters as a probe of the trilinear Higgs boson self-interaction}},  {\em Phys. Rev. D} {\bf 95} (2017), no.~9 093004, [\href{http://arxiv.org/abs/1702.07678}{{\tt arXiv:1702.07678}}].

\bibitem{Degrassi:2017ucl}
G.~Degrassi, M.~Fedele, and P.~P. Giardino, {\it {Constraints on the trilinear Higgs self coupling from precision observables}},  {\em JHEP} {\bf 04} (2017) 155, [\href{http://arxiv.org/abs/1702.01737}{{\tt arXiv:1702.01737}}].

\bibitem{DiVita:2017vrr}
S.~Di~Vita, G.~Durieux, C.~Grojean, J.~Gu, Z.~Liu, G.~Panico, M.~Riembau, and T.~Vantalon, {\it {A global view on the Higgs self-coupling at lepton colliders}},  {\em JHEP} {\bf 02} (2018) 178, [\href{http://arxiv.org/abs/1711.03978}{{\tt arXiv:1711.03978}}].

\bibitem{Bobeth:2015zqa}
C.~Bobeth and U.~Haisch, {\it {Anomalous triple gauge couplings from $B$-meson and kaon observables}},  {\em JHEP} {\bf 09} (2015) 018, [\href{http://arxiv.org/abs/1503.04829}{{\tt arXiv:1503.04829}}].

\bibitem{Falkowski:2015jaa}
A.~Falkowski, M.~Gonzalez-Alonso, A.~Greljo, and D.~Marzocca, {\it {Global constraints on anomalous triple gauge couplings in effective field theory approach}},  {\em Phys. Rev. Lett.} {\bf 116} (2016), no.~1 011801, [\href{http://arxiv.org/abs/1508.00581}{{\tt arXiv:1508.00581}}].

\bibitem{Durieux:2017rsg}
G.~Durieux, C.~Grojean, J.~Gu, and K.~Wang, {\it {The leptonic future of the Higgs}},  {\em JHEP} {\bf 09} (2017) 014, [\href{http://arxiv.org/abs/1704.02333}{{\tt arXiv:1704.02333}}].

\bibitem{Pomarol:2013zra}
A.~Pomarol and F.~Riva, {\it {Towards the Ultimate SM Fit to Close in on Higgs Physics}},  {\em JHEP} {\bf 01} (2014) 151, [\href{http://arxiv.org/abs/1308.2803}{{\tt arXiv:1308.2803}}].

\bibitem{Corbett:2013pja}
T.~Corbett, O.~J.~P. \'Eboli, J.~Gonzalez-Fraile, and M.~C. Gonzalez-Garcia, {\it {Determining Triple Gauge Boson Couplings from Higgs Data}},  {\em Phys. Rev. Lett.} {\bf 111} (2013) 011801, [\href{http://arxiv.org/abs/1304.1151}{{\tt arXiv:1304.1151}}].

\bibitem{Masso:2014xra}
E.~Masso, {\it {An Effective Guide to Beyond the Standard Model Physics}},  {\em JHEP} {\bf 10} (2014) 128, [\href{http://arxiv.org/abs/1406.6376}{{\tt arXiv:1406.6376}}].

\bibitem{Ellis:2014dva}
J.~Ellis, V.~Sanz, and T.~You, {\it {Complete Higgs Sector Constraints on Dimension-6 Operators}},  {\em JHEP} {\bf 07} (2014) 036, [\href{http://arxiv.org/abs/1404.3667}{{\tt arXiv:1404.3667}}].

\bibitem{Ellis:2014jta}
J.~Ellis, V.~Sanz, and T.~You, {\it {The Effective Standard Model after LHC Run I}},  {\em JHEP} {\bf 03} (2015) 157, [\href{http://arxiv.org/abs/1410.7703}{{\tt arXiv:1410.7703}}].

\bibitem{Dumont:2013wma}
B.~Dumont, S.~Fichet, and G.~von Gersdorff, {\it {A Bayesian view of the Higgs sector with higher dimensional operators}},  {\em JHEP} {\bf 07} (2013) 065, [\href{http://arxiv.org/abs/1304.3369}{{\tt arXiv:1304.3369}}].

\bibitem{Dawson:2019clf}
S.~Dawson and P.~P. Giardino, {\it {Electroweak and QCD corrections to $Z$ and $W$ pole observables in the standard model EFT}},  {\em Phys. Rev. D} {\bf 101} (2020), no.~1 013001, [\href{http://arxiv.org/abs/1909.02000}{{\tt arXiv:1909.02000}}].

\bibitem{Haisch:2024wnw}
U.~Haisch and L.~Schnell, {\it {Precision tests of third-generation four-quark operators: matching SMEFT to LEFT}},  \href{http://arxiv.org/abs/2410.13304}{{\tt arXiv:2410.13304}}.

\bibitem{Ge:2024pfn}
S.-F. Ge, Z.~Qian, M.~J. Ramsey-Musolf, and J.~Zhou, {\it {New Physics Off the $Z$-Pole: $e^+ e^- \rightarrow f \bar f$ at Future Lepton Colliders}},  \href{http://arxiv.org/abs/2410.17605}{{\tt arXiv:2410.17605}}.

\bibitem{Haisch:2020ahr}
U.~Haisch, M.~Ruhdorfer, E.~Salvioni, E.~Venturini, and A.~Weiler, {\it {Singlet night in Feynman-ville: one-loop matching of a real scalar}},  {\em JHEP} {\bf 04} (2020) 164, [\href{http://arxiv.org/abs/2003.05936}{{\tt arXiv:2003.05936}}]. [Erratum: JHEP 07, 066 (2020)].

\bibitem{Fuentes-Martin:2022jrf}
J.~Fuentes-Mart\'\i{}n, M.~K\"onig, J.~Pag\`es, A.~E. Thomsen, and F.~Wilsch, {\it {A proof of concept for matchete: an automated tool for matching effective theories}},  {\em Eur. Phys. J. C} {\bf 83} (2023), no.~7 662, [\href{http://arxiv.org/abs/2212.04510}{{\tt arXiv:2212.04510}}].

\bibitem{DiLuzio:2018jwd}
L.~Di~Luzio, R.~Gr\"ober, and G.~Panico, {\it {Probing new electroweak states via precision measurements at the LHC and future colliders}},  {\em JHEP} {\bf 01} (2019) 011, [\href{http://arxiv.org/abs/1810.10993}{{\tt arXiv:1810.10993}}].

\bibitem{Cirelli:2005uq}
M.~Cirelli, N.~Fornengo, and A.~Strumia, {\it {Minimal dark matter}},  {\em Nucl. Phys. B} {\bf 753} (2006) 178--194, [\href{http://arxiv.org/abs/hep-ph/0512090}{{\tt hep-ph/0512090}}].

\bibitem{Harigaya:2015yaa}
K.~Harigaya, K.~Ichikawa, A.~Kundu, S.~Matsumoto, and S.~Shirai, {\it {Indirect Probe of Electroweak-Interacting Particles at Future Lepton Colliders}},  {\em JHEP} {\bf 09} (2015) 105, [\href{http://arxiv.org/abs/1504.03402}{{\tt arXiv:1504.03402}}].

\bibitem{Bottaro:2021snn}
S.~Bottaro, D.~Buttazzo, M.~Costa, R.~Franceschini, P.~Panci, D.~Redigolo, and L.~Vittorio, {\it {Closing the window on WIMP Dark Matter}},  {\em Eur. Phys. J. C} {\bf 82} (2022), no.~1 31, [\href{http://arxiv.org/abs/2107.09688}{{\tt arXiv:2107.09688}}].

\bibitem{Bottaro:2022one}
S.~Bottaro, D.~Buttazzo, M.~Costa, R.~Franceschini, P.~Panci, D.~Redigolo, and L.~Vittorio, {\it {The last complex WIMPs standing}},  {\em Eur. Phys. J. C} {\bf 82} (2022), no.~11 992, [\href{http://arxiv.org/abs/2205.04486}{{\tt arXiv:2205.04486}}].

\bibitem{deBlas:2017xtg}
J.~de~Blas, J.~C. Criado, M.~Perez-Victoria, and J.~Santiago, {\it {Effective description of general extensions of the Standard Model: the complete tree-level dictionary}},  {\em JHEP} {\bf 03} (2018) 109, [\href{http://arxiv.org/abs/1711.10391}{{\tt arXiv:1711.10391}}].

\bibitem{Born:2024mgz}
L.~Born, J.~Fuentes-Mart\'\i{}n, S.~Kvedarait\.{e}, and A.~E. Thomsen, {\it {Two-Loop Running in the Bosonic SMEFT Using Functional Methods}},  \href{http://arxiv.org/abs/2410.07320}{{\tt arXiv:2410.07320}}.

\bibitem{Dedes:2017zog}
A.~Dedes, W.~Materkowska, M.~Paraskevas, J.~Rosiek, and K.~Suxho, {\it {Feynman rules for the Standard Model Effective Field Theory in R$_{\xi}$ -gauges}},  {\em JHEP} {\bf 06} (2017) 143, [\href{http://arxiv.org/abs/1704.03888}{{\tt arXiv:1704.03888}}].

\bibitem{Dedes:2019uzs}
A.~Dedes, M.~Paraskevas, J.~Rosiek, K.~Suxho, and L.~Trifyllis, {\it {SmeftFR \textendash{} Feynman rules generator for the Standard Model Effective Field Theory}},  {\em Comput. Phys. Commun.} {\bf 247} (2020) 106931, [\href{http://arxiv.org/abs/1904.03204}{{\tt arXiv:1904.03204}}].

\bibitem{Dedes:2023zws}
A.~Dedes, J.~Rosiek, M.~Ryczkowski, K.~Suxho, and L.~Trifyllis, {\it {SmeftFR v3 \textendash{} Feynman rules generator for the Standard Model Effective Field Theory}},  {\em Comput. Phys. Commun.} {\bf 294} (2024) 108943, [\href{http://arxiv.org/abs/2302.01353}{{\tt arXiv:2302.01353}}].

\bibitem{Hahn:2000kx}
T.~Hahn, {\it {Generating Feynman diagrams and amplitudes with FeynArts 3}},  {\em Comput. Phys. Commun.} {\bf 140} (2001) 418--431, [\href{http://arxiv.org/abs/hep-ph/0012260}{{\tt hep-ph/0012260}}].

\bibitem{Mertig:1990an}
R.~Mertig, M.~Bohm, and A.~Denner, {\it {FEYN CALC: Computer algebraic calculation of Feynman amplitudes}},  {\em Comput. Phys. Commun.} {\bf 64} (1991) 345--359.

\bibitem{Shtabovenko:2016sxi}
V.~Shtabovenko, R.~Mertig, and F.~Orellana, {\it {New Developments in FeynCalc 9.0}},  {\em Comput. Phys. Commun.} {\bf 207} (2016) 432--444, [\href{http://arxiv.org/abs/1601.01167}{{\tt arXiv:1601.01167}}].

\bibitem{Shtabovenko:2020gxv}
V.~Shtabovenko, R.~Mertig, and F.~Orellana, {\it {FeynCalc 9.3: New features and improvements}},  {\em Comput. Phys. Commun.} {\bf 256} (2020) 107478, [\href{http://arxiv.org/abs/2001.04407}{{\tt arXiv:2001.04407}}].

\bibitem{Shtabovenko:2023idz}
V.~Shtabovenko, R.~Mertig, and F.~Orellana, {\it {FeynCalc 10: Do multiloop integrals dream of computer codes?}},  {\em Comput. Phys. Commun.} {\bf 306} (2025) 109357, [\href{http://arxiv.org/abs/2312.14089}{{\tt arXiv:2312.14089}}].

\bibitem{Aebischer:2017ugx}
J.~Aebischer et~al., {\it {WCxf: an exchange format for Wilson coefficients beyond the Standard Model}},  {\em Comput. Phys. Commun.} {\bf 232} (2018) 71--83, [\href{http://arxiv.org/abs/1712.05298}{{\tt arXiv:1712.05298}}].

\end{thebibliography}\endgroup

\end{document}